\documentclass[pra,showpacs,twocolumn,superscriptaddress]{revtex4}

\usepackage{graphics}
\usepackage{multirow}
\usepackage[usenames]{color}

\def\temp{1.35}%
\let\tempp=\relax
\expandafter\ifx\csname psboxversion\endcsname\relax
\else
    \ifdim\temp cm>\psboxversion cm
    \else
      \let\temp=\psboxversion
      \let\tempp= 
    \fi
\fi
\tempp
\let\psboxversion=\temp
\catcode`\@=11
\def\psfortextures{
\def\PSspeci@l##1##2{%
\special{illustration ##1\space scaled ##2}%
}}%
\def\psfordvitops{
\def\PSspeci@l##1##2{%
\special{dvitops: import ##1\space \the\drawingwd \the\drawinght}%
}}%
\def\psfordvips{
\def\PSspeci@l##1##2{%
\d@my=0.1bp \d@mx=\drawingwd \divide\d@mx by\d@my
\includegraphics{##1\space}}}%
\def\psforoztex{
\def\PSspeci@l##1##2{%
\special{##1 \space
      ##2 1000 div dup scale
      \number-\psllx\space\space \number-\pslly\space\space translate
}}}%
\def\psfordvitps{
\def\dvitpsLiter@ldim##1{\dimen0=##1\relax
\special{dvitps: Literal "\number\dimen0\space"}}%
\def\PSspeci@l##1##2{%
\at(0bp;\drawinght){%
\special{dvitps: Include0 "psfig.psr"}
\dvitpsLiter@ldim{\drawingwd}%
\dvitpsLiter@ldim{\drawinght}%
\dvitpsLiter@ldim{\psllx bp}%
\dvitpsLiter@ldim{\pslly bp}%
\dvitpsLiter@ldim{\psurx bp}%
\dvitpsLiter@ldim{\psury bp}%
\special{dvitps: Literal "startTexFig"}%
\special{dvitps: Include1 "##1"}%
\special{dvitps: Literal "endTexFig"}%
}}}%
\def\psfordvialw{
\def\PSspeci@l##1##2{
\special{language "PostScript",
position = "bottom left",
literal "  \psllx\space \pslly\space translate
  ##2 1000 div dup scale
  -\psllx\space -\pslly\space translate",
include "##1"}
}}%
\def\psforptips{
\def\PSspeci@l##1##2{{
\d@mx=\psurx bp
\advance \d@mx by -\psllx bp
\divide \d@mx by 1000\multiply\d@mx by \xscale
\incm{\d@mx}
\let\tmpx\dimincm
\d@my=\psury bp
\advance \d@my by -\pslly bp
\divide \d@my by 1000\multiply\d@my by \xscale
\incm{\d@my}
\let\tmpy\dimincm
\d@mx=-\psllx bp
\divide \d@mx by 1000\multiply\d@mx by \xscale
\d@my=-\pslly bp
\divide \d@my by 1000\multiply\d@my by \xscale
\at(\d@mx;\d@my){\special{ps:##1 x=\tmpx cm, y=\tmpy cm}}
}}}%
\def\psonlyboxes{
\def\PSspeci@l##1##2{%
\at(0cm;0cm){\boxit{\vbox to\drawinght
  {\vss\hbox to\drawingwd{\at(0cm;0cm){\hbox{({\tt##1})}}\hss}}}}
}}%
\def\psloc@lerr#1{%
\let\savedPSspeci@l=\PSspeci@l%
\def\PSspeci@l##1##2{%
\at(0cm;0cm){\boxit{\vbox to\drawinght
  {\vss\hbox to\drawingwd{\at(0cm;0cm){\hbox{({\tt##1}) #1}}\hss}}}}
\let\PSspeci@l=\savedPSspeci@l
}}%
\newread\pst@mpin
\newdimen\drawinght\newdimen\drawingwd
\newdimen\psxoffset\newdimen\psyoffset
\newbox\drawingBox
\newcount\xscale \newcount\yscale \newdimen\pscm\pscm=1cm
\newdimen\d@mx \newdimen\d@my
\newdimen\pswdincr \newdimen\pshtincr
\let\ps@nnotation=\relax
{\catcode`\|=0 |catcode`|\=12 |catcode`|
|catcode`#=12 |catcode`*=14
|xdef|backslashother{\}*
|xdef|percentother{
|xdef|tildeother{~}*
|xdef|sharpother{#}*
}%
\def\R@moveMeaningHeader#1:->{}%
\def\uncatcode#1{%
\edef#1{\expandafter\R@moveMeaningHeader\meaning#1}}%
\def\execute#1{#1}
\def\psm@keother#1{\catcode`#112\relax}
\def\executeinspecs#1{%
\execute{\begingroup\let\do\psm@keother\dospecials\catcode`\^^M=9#1\endgroup}}%
\def\@mpty{}%
\def\matchexpin#1#2{
  \fi%
  \edef\tmpb{{#2}}%
  \expandafter\makem@tchtmp\tmpb%
  \edef\tmpa{#1}\edef\tmpb{#2}%
  \expandafter\expandafter\expandafter\m@tchtmp\expandafter\tmpa\tmpb\endm@tch%
  \if\match%
}%
\def\matchin#1#2{%
  \fi%
  \makem@tchtmp{#2}%
  \m@tchtmp#1#2\endm@tch%
  \if\match%
}%
\def\makem@tchtmp#1{\def\m@tchtmp##1#1##2\endm@tch{%
  \def\tmpa{##1}\def\tmpb{##2}\let\m@tchtmp=\relax%
  \ifx\tmpb\@mpty\def\match{YN}%
  \else\def\match{YY}\fi%
}}%
\def\incm#1{{\psxoffset=1cm\d@my=#1
 \d@mx=\d@my
  \divide\d@mx by \psxoffset
  \xdef\dimincm{\number\d@mx.}
  \advance\d@my by -\number\d@mx cm
  \multiply\d@my by 100
 \d@mx=\d@my
  \divide\d@mx by \psxoffset
  \edef\dimincm{\dimincm\number\d@mx}
  \advance\d@my by -\number\d@mx cm
  \multiply\d@my by 100
 \d@mx=\d@my
  \divide\d@mx by \psxoffset
  \xdef\dimincm{\dimincm\number\d@mx}
}}%
\newif\ifNotB@undingBox
\newhelp\PShelp{Proceed: you'll have a 5cm square blank box instead of
your graphics.}%
\def\s@tsize#1 #2 #3 #4\@ndsize{
  \def\psllx{#1}\def\pslly{#2}%
  \def\psurx{#3}\def\psury{#4}
  \ifx\psurx\@mpty\NotB@undingBoxtrue
  \else
    \drawinght=#4bp\advance\drawinght by-#2bp
    \drawingwd=#3bp\advance\drawingwd by-#1bp
  \fi
  }%
\def\sc@nBBline#1:#2\@ndBBline{\edef\p@rameter{#1}\edef\v@lue{#2}}%
\def\g@bblefirstblank#1#2:{\ifx#1 \else#1\fi#2}%
{\catcode`\%=12
\xdef\B@undingBox{
\def\ReadPSize#1{
 \readfilename#1\relax
 \let\PSfilename=\lastreadfilename
 \openin\pst@mpin=#1\relax
 \ifeof\pst@mpin \errhelp=\PShelp
   \errmessage{I haven't found your postscript file (\PSfilename)}%
   \psloc@lerr{was not found}%
   \s@tsize 0 0 142 142\@ndsize
   \closein\pst@mpin
 \else
   \if\matchexpin{\GlobalInputList}{, \lastreadfilename}%
   \else\xdef\GlobalInputList{\GlobalInputList, \lastreadfilename}%
     \immediate\write\psbj@inaux{\lastreadfilename,}%
   \fi%
   \loop
     \executeinspecs{\catcode`\ =10\global\read\pst@mpin to\n@xtline}%
     \ifeof\pst@mpin
       \errhelp=\PShelp
       \errmessage{(\PSfilename) is not an Encapsulated PostScript File:
           I could not find any \B@undingBox: line.}%
       \edef\v@lue{0 0 142 142:}%
       \psloc@lerr{is not an EPSFile}%
       \NotB@undingBoxfalse
     \else
       \expandafter\sc@nBBline\n@xtline:\@ndBBline
       \ifx\p@rameter\B@undingBox\NotB@undingBoxfalse
         \edef\t@mp{%
           \expandafter\g@bblefirstblank\v@lue\space\space\space}%
         \expandafter\s@tsize\t@mp\@ndsize
       \else\NotB@undingBoxtrue
       \fi
     \fi
   \ifNotB@undingBox\repeat
   \closein\pst@mpin
 \fi
\message{#1}%
}%
\def\psboxto(#1;#2)#3{\vbox{%
   \ReadPSize{#3}%
   \advance\pswdincr by \drawingwd
   \advance\pshtincr by \drawinght
   \divide\pswdincr by 1000
   \divide\pshtincr by 1000
   \d@mx=#1
   \ifdim\d@mx=0pt\xscale=1000
         \else \xscale=\d@mx \divide \xscale by \pswdincr\fi
   \d@my=#2
   \ifdim\d@my=0pt\yscale=1000
         \else \yscale=\d@my \divide \yscale by \pshtincr\fi
   \ifnum\yscale=1000
         \else\ifnum\xscale=1000\xscale=\yscale
                    \else\ifnum\yscale<\xscale\xscale=\yscale\fi
              \fi
   \fi
   \divide\drawingwd by1000 \multiply\drawingwd by\xscale
   \divide\drawinght by1000 \multiply\drawinght by\xscale
   \divide\psxoffset by1000 \multiply\psxoffset by\xscale
   \divide\psyoffset by1000 \multiply\psyoffset by\xscale
   \global\divide\pscm by 1000
   \global\multiply\pscm by\xscale
   \multiply\pswdincr by\xscale \multiply\pshtincr by\xscale
   \ifdim\d@mx=0pt\d@mx=\pswdincr\fi
   \ifdim\d@my=0pt\d@my=\pshtincr\fi
   \message{scaled \the\xscale}%
 \hbox to\d@mx{\hss\vbox to\d@my{\vss
   \global\setbox\drawingBox=\hbox to 0pt{\kern\psxoffset\vbox to 0pt{%
      \kern-\psyoffset
      \PSspeci@l{\PSfilename}{\the\xscale}%
      \vss}\hss\ps@nnotation}%
   \global\wd\drawingBox=\the\pswdincr
   \global\ht\drawingBox=\the\pshtincr
   \global\drawingwd=\pswdincr
   \global\drawinght=\pshtincr
   \baselineskip=0pt
   \copy\drawingBox
 \vss}\hss}%
  \global\psxoffset=0pt
  \global\psyoffset=0pt
  \global\pswdincr=0pt
  \global\pshtincr=0pt 
  \global\pscm=1cm 
}}%
\def\psboxscaled#1#2{\vbox{%
  \ReadPSize{#2}%
  \xscale=#1
  \message{scaled \the\xscale}%
  \divide\pswdincr by 1000 \multiply\pswdincr by \xscale
  \divide\pshtincr by 1000 \multiply\pshtincr by \xscale
  \divide\psxoffset by1000 \multiply\psxoffset by\xscale
  \divide\psyoffset by1000 \multiply\psyoffset by\xscale
  \divide\drawingwd by1000 \multiply\drawingwd by\xscale
  \divide\drawinght by1000 \multiply\drawinght by\xscale
  \global\divide\pscm by 1000
  \global\multiply\pscm by\xscale
  \global\setbox\drawingBox=\hbox to 0pt{\kern\psxoffset\vbox to 0pt{%
     \kern-\psyoffset
     \PSspeci@l{\PSfilename}{\the\xscale}%
     \vss}\hss\ps@nnotation}%
  \advance\pswdincr by \drawingwd
  \advance\pshtincr by \drawinght
  \global\wd\drawingBox=\the\pswdincr
  \global\ht\drawingBox=\the\pshtincr
  \global\drawingwd=\pswdincr
  \global\drawinght=\pshtincr
  \baselineskip=0pt
  \copy\drawingBox
  \global\psxoffset=0pt
  \global\psyoffset=0pt
  \global\pswdincr=0pt
  \global\pshtincr=0pt 
  \global\pscm=1cm
}}%
\def\psbox#1{\psboxscaled{1000}{#1}}%
\newif\ifn@teof\n@teoftrue
\newif\ifc@ntrolline
\newif\ifmatch
\newread\j@insplitin
\newwrite\j@insplitout
\newwrite\psbj@inaux
\immediate\openout\psbj@inaux=psbjoin.aux
\immediate\write\psbj@inaux{\string\joinfiles}%
\immediate\write\psbj@inaux{\jobname,}%
\def\toother#1{\ifcat\relax#1\else\expandafter%
  \toother@ux\meaning#1\endtoother@ux\fi}%
\def\toother@ux#1 #2#3\endtoother@ux{\def\tmp{#3}%
  \ifx\tmp\@mpty\def\tmp{#2}\let\next=\relax%
  \else\def\next{\toother@ux#2#3\endtoother@ux}\fi%
\next}%
\let\readfilenamehook=\relax
\def\re@d{\expandafter\re@daux}
\def\re@daux{\futurelet\nextchar\stopre@dtest}%
\def\re@dnext{\xdef\lastreadfilename{\lastreadfilename\nextchar}%
  \afterassignment\re@d\let\nextchar}%
\def\stopre@d{\egroup\readfilenamehook}%
\def\stopre@dtest{%
  \ifcat\nextchar\relax\let\nextread\stopre@d
  \else
    \ifcat\nextchar\space\def\nextread{%
      \afterassignment\stopre@d\chardef\nextchar=`}%
    \else\let\nextread=\re@dnext
      \toother\nextchar
      \edef\nextchar{\tmp}%
    \fi
  \fi\nextread}%
\def\readfilename{\bgroup%
  \let\\=\backslashother \let\%=\percentother \let\~=\tildeother
  \let\#=\sharpother \xdef\lastreadfilename{}%
  \re@d}%
\xdef\GlobalInputList{\jobname}%
\def\psnewinput{%
  \def\readfilenamehook{
    \if\matchexpin{\GlobalInputList}{, \lastreadfilename}%
    \else\xdef\GlobalInputList{\GlobalInputList, \lastreadfilename}%
      \immediate\write\psbj@inaux{\lastreadfilename,}%
    \fi%
    \let\readfilenamehook=\relax%
    \ps@ldinput\lastreadfilename\relax%
  }\readfilename%
}%
\expandafter\ifx\csname @@input\endcsname\relax    
  \immediate\let\ps@ldinput=\input\def\input{\psnewinput}%
\else
  \immediate\let\ps@ldinput=\@@input
  \def\@@input{\psnewinput}%
\fi%
\def\nowarnopenout{%
 \def\warnopenout##1##2{%
   \readfilename##2\relax
   \message{\lastreadfilename}%
   \immediate\openout##1=\lastreadfilename\relax}}%
\def\warnopenout#1#2{%
 \readfilename#2\relax
 \def\t@mp{TrashMe,psbjoin.aux,psbjoint.tex,}\uncatcode\t@mp
 \if\matchexpin{\t@mp}{\lastreadfilename,}%
 \else
   \immediate\openin\pst@mpin=\lastreadfilename\relax
   \ifeof\pst@mpin
     \else
     \edef\tmp{{If the content of this file is precious to you, this
is your last chance to abort (ie press x or e) and rename it before
retexing (\jobname). If you're sure there's no file
(\lastreadfilename) in the directory of (\jobname), then go on: I'm
simply worried because you have another (\lastreadfilename) in some
directory I'm looking in for inputs...}}%
     \errhelp=\tmp
     \errmessage{I may be about to replace your file named \lastreadfilename}%
   \fi
   \immediate\closein\pst@mpin
 \fi
 \message{\lastreadfilename}%
 \immediate\openout#1=\lastreadfilename\relax}%
{\catcode`\%=12\catcode`\*=14
\gdef\splitfile#1{*
 \readfilename#1\relax
 \immediate\openin\j@insplitin=\lastreadfilename\relax
 \ifeof\j@insplitin
   \message{! I couldn't find and split \lastreadfilename!}*
 \else
   \immediate\openout\j@insplitout=TrashMe
   \message{< Splitting \lastreadfilename\space into}*
   \loop
     \ifeof\j@insplitin
       \immediate\closein\j@insplitin\n@teoffalse
     \else
       \n@teoftrue
       \executeinspecs{\global\read\j@insplitin to\spl@tinline\expandafter
         \ch@ckbeginnewfile\spl@tinline
       \ifc@ntrolline
       \else
         \toks0=\expandafter{\spl@tinline}*
         \immediate\write\j@insplitout{\the\toks0}*
       \fi
     \fi
   \ifn@teof\repeat
   \immediate\closeout\j@insplitout
 \fi\message{>}*
}*
\gdef\ch@ckbeginnewfile#1
 \def\t@mp{#1}*
 \ifx\@mpty\t@mp
   \def\t@mp{#3}*
   \ifx\@mpty\t@mp
     \global\c@ntrollinefalse
   \else
     \immediate\closeout\j@insplitout
     \warnopenout\j@insplitout{#2}*
     \global\c@ntrollinetrue
   \fi
 \else
   \global\c@ntrollinefalse
 \fi}*
\gdef\joinfiles#1\into#2{*
 \message{< Joining following files into}*
 \warnopenout\j@insplitout{#2}*
 \message{:}*
 {*
 \edef\w@##1{\immediate\write\j@insplitout{##1}}*
\w@{
\w@{
\w@{
\w@{
\w@{
\w@{
\w@{
\w@{
\w@{
\w@{
\w@{\string\input\space psbox.tex}*
\w@{\string\splitfile{\string\jobname}}*
\w@{\string\let\string\autojoin=\string\relax}*
}*
 \expandafter\tre@tfilelist#1, \endtre@t
 \immediate\closeout\j@insplitout
 \message{>}*
}*
\gdef\tre@tfilelist#1, #2\endtre@t{*
 \readfilename#1\relax
 \ifx\@mpty\lastreadfilename
 \else
   \immediate\openin\j@insplitin=\lastreadfilename\relax
   \ifeof\j@insplitin
     \errmessage{I couldn't find file \lastreadfilename}*
   \else
     \message{\lastreadfilename}*
     \immediate\write\j@insplitout{
     \executeinspecs{\global\read\j@insplitin to\oldj@ininline}*
     \loop
       \ifeof\j@insplitin\immediate\closein\j@insplitin\n@teoffalse
       \else\n@teoftrue
         \executeinspecs{\global\read\j@insplitin to\j@ininline}*
         \toks0=\expandafter{\oldj@ininline}*
         \let\oldj@ininline=\j@ininline
         \immediate\write\j@insplitout{\the\toks0}*
       \fi
     \ifn@teof
     \repeat
   \immediate\closein\j@insplitin
   \fi
   \tre@tfilelist#2, \endtre@t
 \fi}*
}%
\def\autojoin{%
 \immediate\write\psbj@inaux{\string\into{psbjoint.tex}}%
 \immediate\closeout\psbj@inaux
 \expandafter\joinfiles\GlobalInputList\into{psbjoint.tex}%
}%
\def\centinsert#1{\midinsert\line{\hss#1\hss}\endinsert}%
\def\psannotate#1#2{\vbox{%
  \def\ps@nnotation{#2\global\let\ps@nnotation=\relax}#1}}%
\def\pscaption#1#2{\vbox{%
   \setbox\drawingBox=#1
   \copy\drawingBox
   \vskip\baselineskip
   \vbox{\hsize=\wd\drawingBox\setbox0=\hbox{#2}%
     \ifdim\wd0>\hsize
       \noindent\unhbox0\tolerance=5000
    \else\centerline{\box0}%
    \fi
}}}%
\def\at(#1;#2)#3{\setbox0=\hbox{#3}\ht0=0pt\dp0=0pt
  \rlap{\kern#1\vbox to0pt{\kern-#2\box0\vss}}}%
\newdimen\gridht \newdimen\gridwd
\def\gridfill(#1;#2){%
  \setbox0=\hbox to 1\pscm
  {\vrule height1\pscm width.4pt\leaders\hrule\hfill}%
  \gridht=#1
  \divide\gridht by \ht0
  \multiply\gridht by \ht0
  \gridwd=#2
  \divide\gridwd by \wd0
  \multiply\gridwd by \wd0
  \advance \gridwd by \wd0
  \vbox to \gridht{\leaders\hbox to\gridwd{\leaders\box0\hfill}\vfill}}%
\def\fillinggrid{\at(0cm;0cm){\vbox{%
  \gridfill(\drawinght;\drawingwd)}}}%
\def\textleftof#1:{%
  \setbox1=#1
  \setbox0=\vbox\bgroup
    \advance\hsize by -\wd1 \advance\hsize by -2em}%
\def\textrightof#1:{%
  \setbox0=#1
  \setbox1=\vbox\bgroup
    \advance\hsize by -\wd0 \advance\hsize by -2em}%
\def\endtext{%
  \egroup
  \hbox to \hsize{\valign{\vfil##\vfil\cr%
\box0\cr%
\noalign{\hss}\box1\cr}}}%
\def\frameit#1#2#3{\hbox{\vrule width#1\vbox{%
  \hrule height#1\vskip#2\hbox{\hskip#2\vbox{#3}\hskip#2}%
        \vskip#2\hrule height#1}\vrule width#1}}%
\def\boxit#1{\frameit{0.4pt}{0pt}{#1}}%
\catcode`\@=12 
\psfordvips   

\newcommand {\mb}[1]{\mbox{\boldmath{${#1}$}}}

\begin{document}

\title{Momentum--space engineering of gaseous Bose--Einstein
condensates}
\author{Mark Edwards}
\affiliation{Department of Physics, Georgia Southern University,
Statesboro, GA 30460--8031 USA}
\affiliation{National Institute of Standards and Technology, 
Gaithersburg, MD 20899, USA}
\author{Brandon Benton}
\affiliation{Department of Physics, Georgia Southern University,
Statesboro, GA 30460--8031 USA}
\author{Jeffrey Heward}
\affiliation{Department of Physics, Georgia Southern University,
Statesboro, GA 30460--8031 USA}
\author{Charles W.\ Clark}
\affiliation{Joint Quantum Insitute, National Institute of Standards 
and Technology and the University of Maryland, Gaithersburg, MD 20899, USA}

\date{\today}

\begin{abstract}
We show how the momentum distribution of gaseous Bose--Einstein
condensates can be shaped by applying a sequence of standing--wave 
laser pulses.  We present a theory, whose validity for 
was demonstrated in an earlier experiment [L.\ Deng, et al., 
\prl {\bf 83}, 5407 (1999)], of the effect of a two--pulse sequence on 
the condensate wavefunction in momentum space. We generalize the previous 
result to the case of $N$ pulses of arbitrary intensity separated by 
arbitrary intervals and show how these parameters can be engineered to 
produce a desired final momentum distribution.  We find that several 
momentum distributions, important in atom--interferometry applications, 
can be engineered with high fidelity with two or three pulses.
\end{abstract}


\maketitle

\section{Introduction}
\label{intro}

The ability to create gaseous Bose--Einstein condensates in the 
laboratory~\cite{Science.269.198,PhysRevLett.75.1687,
PhysRevLett.75.3969,Pethick_and_Smith,Pitaevskii_and_Stringari}
has led to new vistas in the field of atom interferometry. This is
particularly true when laser light is used to manipulate atoms to produce 
matter--wave interference patterns.  Atom interferometers using light 
gratings acting on matter waves have been used in a variety of fundamental 
studies such as how large a composite object can display interference 
effects~\cite{PhysRevLett.74.4783}, decoherence 
studies~\cite{PhysRevLett.91.090408,AngewChemIntEd.47.6195}, origins of 
phase shifts under various circumstances, properties of Bose--Einstein 
condensates~\cite{PhysRevLett.78.582,PhysRevLett.82.3008,PhysRevLett.83.5407}, 
and testing the charge neutrality of atoms~\cite{PhysRevA.47.4663}.  
Atom interferometers are also at the heart of a host of practical devices 
used for making precision measurements.  These include gravimeters, 
gyroscopes, and gradiometers all of which have important applications 
in precision navigation
~\cite{Metrologia.38.25,PhysRevA.65.033608,ClassQuantumGrav.17.2385}.  
Such interferometers also have applications in atomic physics such as 
atomic polarizability measurements and Casimir--Polder potentials for 
atoms near surfaces~\cite{EurPhysJD.38.353}.  More in--depth information 
about the uses of atom interferometry can be found in 
Ref.\ \cite{RevModPhys.81.1051}.

One of the crucial elements of an atom interferometer is initial
state selection of the atoms and these states are generally states
of localized momentum~\cite{RevModPhys.81.1051}.  Momentum--state
selection techniques are quite varied and can range from using a pair
of collimating slits that select thermal atoms with limited transverse
momentum to Bose--Einstein condensation of atom clouds via techniques
that are a combination of laser cooling and trapping and evaporative
cooling. In this paper we show how applying a sequence of short--time,
standing--wave laser pulses to a Bose--Einstein condensate (BEC) can
be used as a tool for the state selection step of a atom interferometric
experiment.  We will also see that preparation of several important
classes of momentum states can be achieved through the application of
just two or three pulses.

Previously a sequence of two short--time, standing--wave 
pulses (sometimes called Kapitza--Dirac pulses) has been used as the beam
splitter in the experimental realization of a Michelson atom interferometer 
for a BEC formed on an atom chip~\cite{PhysRevLett.94.090405}.  Optimization
of the two--pulse sequence was determined by studying a two--state
truncation of the Raman--Nath equations~\cite{PhysRevA.71.043602}.
Other related beam--splitter techniques which produce specific momentum
orders~\cite{PhysRevA.58.4801,PhysRevA.55.4382,PhysRevA.68.023610} 
have been studied.  However, these techniques represent a different
strategy~\cite{PhysRevA.71.043602} in the standing--wave control of 
atomic motion.

This paper is organized as follows.  In Section \ref{two-pulse_theory},
we derive a theory for how the condensate wave function is changed 
following the application of two pulses of different intensity and
separated by a time interval on the order of the Talbot time.  The
Talbot time is $T_{T}=h/E_{recoil}$ where $E_{recoil}$ is the recoil
energy of the atom for the laser light used to make the standing--wave
pulses.  This theory is composed of two parts, (1) the effect of a
pulse on the condensate wave function and (2) evolution of the wave
function between pulses.  These two elements are used to follow the
steps of a two--pulse sequence to derive an expression for the amplitude
for atoms to jump from the zero--momentum initial state to an arbitrary
momentum order.  In Section \ref{general}, we derive two general 
symmetries of the two--pulse amplitude and some values for special
time interval values.  This section also presents a physical interpretation
of the amplitude.  This interpretation is especially useful for 
generalizing the two--pulse result.  Section \ref{n-pulse_theory}
contains a derivation of the general $N$--pulse amplitude.  Section
\ref{eng_dist} presents the least--squares method for designing a
general sequence of $N$ pulses which gives a specified momentum 
probability distribution.  Examples of two--pulse and three--pulse
sequences are given for some important momentum distributions.
Conclusions are presented in Section \ref{conclusion}.

\section{Two--Pulse Theory}
\label{two-pulse_theory}

In this section we derive how the condensate wave function is changed 
when two short--time, standing--wave laser pulses are applied. We 
assume that each pulse is square--shaped in time and that its duration, 
$\delta t$, is short enough such that there is no
appreciable spontaneous emission during the pulse and short enough so that
are no effects of atom--atom interactions during the pulse. This will 
happen if $\delta t < \hbar/\Gamma$ where $\Gamma$ is the natural line width 
of the excited state. We also assume that the entire pulse sequence is short 
compared to $\hbar/\mu$, where $\mu$ is the chemical potential, i.e., short 
enough so that there are no effects due to the interaction during the pulse 
sequence.  It is important to note that, under these assumptions, many--body
effects can be neglected.  

We assume that the first pulse is applied at time $t_{1}$ and 
has duration $\delta t_{1}$ and that the second pulse is applied at time 
$t_{2}$ and has duration $\delta t_{2}$.  We also assume that the time 
interval, $t_{2}-t_{1}$, is small enough so that the maximum distance that 
condensate atoms in non--zero momentum states move after the first pulse 
is small compared to their de Broglie wavelengths.  This is the 
{\em Raman--Nath} regime.
  
Our approach will be to use these conditions to approximate the effect 
that a single pulse has on the condensate wave function and separately to 
approximate how the wave function evolves between pulses.  With these 
effects in hand, we can then follow the steps of the pulse sequence to 
determine the overall effect of the full pulse sequence on the wave 
function.

\subsection{Effect of the first pulse}
\label{first_pulse}

Consider atoms in a Bose--Einstein condensate interacting with a 
pair of linearly polarized, counterpropagating laser pulses.  Each atom is 
modeled as a two--level system having a ground state $|g\rangle$ and an excited
state $|e\rangle$.  We denote the position of the atom's center--of--mass 
(CM) relative to an arbitrary coordinate system by ${\bf r}_{a}$ and the 
position of the atomic electron relative to the CM by ${\bf r}_{e}$ so that 
the position vector of the electron in the arbitrary system is 
${\bf r}_{a} + {\bf r}_{e}$.  

The Hamiltonian for a single atom interacting with the light and including 
the CM motion is given by
\begin{eqnarray}
H &=& 
H_{0}^{(g)}\left({\bf r}_{a}\right)
\left|g\rangle\langle g\right| +
H_{0}^{(e)}\left({\bf r}_{a}\right)
\left|e\rangle\langle e\right|\nonumber\\
&& 
+\,
E_{g}\left|g\rangle\langle g\right| +
E_{e}\left|e\rangle\langle e\right| + 
V_{{\rm laser}}\left({\bf r}_{a},t\right).
\end{eqnarray}
Where
\begin{equation}
H_{0}^{(k)}\left({\bf r}_{a}\right) = 
\frac{{\bf p}_{a}^{2}}{2m} + 
V_{{\rm trap}}^{(k)}\left({\bf r}_{a}\right),\quad
k = g, e
\end{equation}
are the energies associated with the CM motion of the atom in the ground 
and excited states.  The difference in trap potentials derives from the 
different magnetic moments for the two internal states.  Both potentials 
are assumed to be harmonic here.

The term $V_{{\rm laser}}$ is the usual dipole interaction and is written 
as
\begin{eqnarray}
V_{{\rm laser}}\left({\bf r}_{a},t\right) &=&
2\hbar\Omega_{0}\cos\left({\bf k}_{L}\cdot{\bf r}_{a}\right)
f\left(t\right)\cos\left(\omega_{L}t\right)\nonumber\\
&&
\times
\left(
\left|e\rangle\langle g\right| +
\left|g\rangle\langle e\right|
\right),
\label{v_laser}
\end{eqnarray}
where the standing--wave laser field is assumed to have the form
\begin{equation}
{\bf E}\left({\bf r}_{a},t\right) = 
2{\cal E}_{0}f\left(t\right)
\hat{{\bf e}}_{L}\cos\left({\bf k}_{L}\cdot{\bf r}_{a}\right)
\cos\left(\omega_{L}t\right).
\label{laser}
\end{equation}
The laser--field amplitude, frequency, and wavevector are denoted by 
${\cal E}_{0}$, $\omega_{L}$, and ${\bf k}_{L}$, respectively, while 
$\hat{{\bf e}}_{L}$ is the laser polarization vector.  The laser--field 
amplitude envelope, $f\left(t\right)$, is assumed to be a rectangular 
pulse centered at an arbitrary time $t_{1}$ with width $\delta t_{1}$.  
The factor $\Omega_{0}$ in Eq.\ (\ref{v_laser}) is the single--photon Rabi 
frequency given by
\begin{equation}
\hbar\Omega_{0} = e{\cal E}_{0}
\langle e\left|{\bf r}_{e}\cdot\hat{{\bf e}}_{L}\right|g\rangle,
\end{equation}
where $e$ is the electron charge.

The condensate orbital is represented by a spinor wave function of the 
form
\begin{eqnarray}
\Psi\left({\bf r}_{a},t\right) &=& 
\left(
\begin{array}{c}
\psi_{g}\left({\bf r}_{a},t\right)\\
\psi_{e}\left({\bf r}_{a},t\right)
\end{array}
\right)
\nonumber\\
&=& 
\psi_{g}\left({\bf r}_{a},t\right)
\left|g\right\rangle +
\psi_{e}\left({\bf r}_{a},t\right)
\left|e\right\rangle.
\end{eqnarray}
These components satisfy the multicomponent Gross--Pitaevskii (GP)
equations
\begin{eqnarray}
i\hbar\frac{\partial\psi_{g}}{\partial t} &=&
H_{0}^{(g)}\psi_{g}\left({\bf r}_{a},t\right) + 
E_{g}\psi_{g}\left({\bf r}_{a},t\right)\nonumber\\ 
&+& 
2\hbar\Omega_{0}\cos\left({\bf k}_{L}\cdot{\bf r}_{a}\right)
f\left(t\right)\cos\left(\omega_{L}t\right)
\psi_{e}\left({\bf r}_{a},t\right)\nonumber\\
&+& 
gN\left(
\left|\psi_{g}\left({\bf r}_{a},t\right)\right|^{2} +
\left|\psi_{e}\left({\bf r}_{a},t\right)\right|^{2}
\right)\psi_{g}\left({\bf r}_{a},t\right),\nonumber\\
\label{mcgp_g}
\end{eqnarray}
and
\begin{eqnarray}
i\hbar\frac{\partial\psi_{e}}{\partial t} &=&
H_{0}^{(e)}\psi_{e}\left({\bf r}_{a},t\right) + 
E_{e}\psi_{e}\left({\bf r}_{a},t\right)\nonumber\\ 
&+& 
2\hbar\Omega_{0}\cos\left({\bf k}_{L}\cdot{\bf r}_{a}\right)
f\left(t\right)\cos\left(\omega_{L}t\right)
\psi_{g}\left({\bf r}_{a},t\right)\nonumber\\
&+& 
gN\left(
\left|\psi_{g}\left({\bf r}_{a},t\right)\right|^{2} +
\left|\psi_{e}\left({\bf r}_{a},t\right)\right|^{2}
\right)\psi_{e}\left({\bf r}_{a},t\right).\nonumber\\
\label{mcgp_e}
\end{eqnarray}
The solution of these equations, over the time of the laser pulse 
can be approximated if (1) the pulse time is short enough so that 
$\delta t_{1} \ll \hbar/\mu$ where $\mu$ is the condensate chemical 
potential, and (2) if the single--photon Rabi frequency is small 
compared to the detuning from resonance, i.e., $\Omega_{0} \ll \Delta$.
The detuning is defined by $\hbar\Delta = E_{e}-E_{g}-\hbar\omega_{L}$.  
We will see that the solution is valid even for a strong pulse.

By approximately solving the multi--component GP equations over the
duration of the pulse, the details of which are given in Appendix \ref{app1}, 
the effect of a short--time, strong--field, standing--wave laser pulse 
on the condensate wave function is given by
\begin{eqnarray}
\phi_{g}\left({\bf r}_{a},t_{1+}\right) &\approx&
e^{i\Omega_{2}\delta t_{1}}\nonumber\\
&&
\times
e^{i\Omega_{2}\delta t_{1}\cos\left(2{\bf k}_{L}\cdot{\bf r}_{a}\right)}
\phi_{g}\left({\bf r}_{a},t_{1-}\right).
\label{1st_pulse}
\end{eqnarray}
where
\begin{equation}
\Omega_{2} \equiv \frac{\Omega_{0}^{2}}{2\Delta}
\end{equation}
is the two--photon Rabi frequency.  

It is possible to represent the effect of the pulse in momentum space
by using the Bessel generating function~\cite{NIST_DLMF}.
\begin{equation}
e^{\frac{1}{2}z\left(t - \frac{1}{t}\right)} =
\sum_{n=-\infty}^{\infty}\,t^{n}J_{n}\left(z\right).
\end{equation}
Letting $t=ie^{2i{\bf k}_{L}\cdot{\bf r}_{a}}$ and 
$z = \Omega_{2}\delta t_{1}$ gives
\begin{equation}
e^{i\Omega_{2}\delta t_{1}\cos\left(2{\bf k}_{L}\cdot{\bf r}_{a}\right)} =
\sum_{n=-\infty}^{\infty}\,
i^{n}J_{n}\left(\Omega_{2}\delta t_{1}\right)
e^{2ni{\bf k}_{L}\cdot{\bf r}_{a}}.
\end{equation}
Hence, the wave function just after the pulse can be written as
\begin{eqnarray}
\phi_{g}\left({\bf r}_{a},t_{1+}\right) &\approx&
e^{i\Omega_{2}\delta t_{1}}\nonumber\\
&\times&
\sum_{n=-\infty}^{\infty}\,
i^{n}J_{n}\left(\Omega_{2}\delta t_{1}\right)
e^{2ni{\bf k}_{L}\cdot{\bf r}_{a}}
\phi_{g}\left({\bf r}_{a},t_{1-}\right).\nonumber\\
\end{eqnarray}
The $e^{2ni{\bf k}_{L}\cdot{\bf r}_{a}}\phi_{g}\left({\bf r}_{a},
t_{1-}\right)$ factor in the $n^{{\rm th}}$ term in the above sum is 
the wave function of the original zero--momentum condensate kicked 
into a momentum state centered at ${\bf p}_{n} = 2n\hbar{\bf k}_{L}$
and expressed in the position--space representation.  Thus the amplitude 
for a condensate atom starting from zero momentum and kicked into momentum 
state $2n\hbar{\bf k}_{L}$ by the pulse is $i^{n}J_{n}\left(\Omega_{2}
\delta t_{1}\right)$.  We now turn to the evolution of the condensate 
wave function between pulses.

\subsection{Evolution between pulses}
\label{between_pulses}

Between the first laser pulse at $t=t_{1}$ and the second at $t=t_{2}$, 
the evolution of an atom in momentum state ${\bf p}_{n} = 2n\hbar{\bf k}_{L}$ 
can be approximated as a free particle under conditions described below.
The phase of such an atom thus evolves as $e^{-iE_{n}(t_{2}-t_{1})/\hbar}$ 
where $E_{n} = p_{n}^{2}/2m$ is its kinetic energy and the atom moves with 
velocity ${\bf v}_{n} = 2n\hbar{\bf k}_{L}/m$.  For non--zero momentum 
states, as long as the number of atoms outcoupled into them is too small to 
be detected experimentally or the time scale over which they evolve is 
small compared to $\hbar/\mu$, mean--field effects can be ignored.  Thus 
we can write the condensate wave function during the time $t_{1}< t < t_{2}$ 
as
\begin{eqnarray}
\phi_{g}\left({\bf r}_{a},t\right) &=&
e^{i\Omega_{2}\delta t_{1}}
\sum_{n=-\infty}^{\infty}\,
i^{n}J_{n}\left(\Omega_{2}\delta t\right)
e^{-iE_{n}\left(t-t_{1}\right)/\hbar}\nonumber\\
&\times&
e^{2ni{\bf k}_{L}\cdot{\bf r}_{a}}
\phi_{g}\left({\bf r}_{a}-{\bf v}_{n}\left(t-t_{1}\right),t_{1}\right)
\label{free_evolution}
\end{eqnarray}

It is clear that, although the above equation implies that there is a 
finite probability for atoms to be outcoupled into a momentum state that 
is {\em any} multiple of $2\hbar{\bf k}_{L}$, above some maximum order, 
$n_{max}$, there will be too few atoms present to be detected experimentally.
Here we will assume that the {\em Raman--Nath} approximation holds, that is, 
the only momentum orders appreciably populated are ones in which the atoms 
moved only a small fraction of their de Broglie wavelengths during the 
time between pulses.  In this case,
\begin{eqnarray}
\phi_{g}\left({\bf r}_{a},t\right) &\approx&
e^{i\Omega_{2}\delta t_{1}}
\sum_{n=-n_{max}}^{n_{max}}\,
i^{n}J_{n}\left(\Omega_{2}\delta t_{1}\right)
e^{-iE_{n}\left(t-t_{1}\right)/\hbar}\nonumber\\
&\times&
e^{2ni{\bf k}_{L}\cdot{\bf r}_{a}}
\phi_{g}\left({\bf r}_{a},t_{1}\right).
\label{raman_nath}
\end{eqnarray}
Note that we have neglected the motion of the non--zero momentum orders 
over the time interval $t-t_{1}$.  We shall assume that the Raman--Nath 
approximation holds hereafter.  Later, when we consider the $N$--pulse 
case, we will assume that this holds for the entire pulse sequence.  Next 
we analyze the effect of the second pulse.

\subsection{Effect of the second pulse}
\label{second_pulse}

The effect of the second pulse can be described by applying
the exponential in Eq.\ (\ref{1st_pulse}) to the wave function
in the Eq.\ (\ref{raman_nath}).  The wave function just after
the application of the second pulse is
\begin{eqnarray}
\phi_{g}\left({\bf r}_{a},t_{2+}\right) &\approx&
e^{i\Omega_{2}(\delta t_{1}+\delta t_{2})}\nonumber\\
&\times&
\sum_{n=-n_{max}}^{n_{max}}\,
i^{n}J_{n}\left(\Omega_{2}\delta t_{1}\right)
e^{-iE_{n}(t_{2}-t_{1})/\hbar}\nonumber\\
&\times&
e^{2ni{\bf k}_{L}\cdot{\bf r}_{a}}
e^{i\Omega_{2}\delta t_{2}\cos\left(2{\bf k}_{L}\cdot{\bf r}_{a}\right)}
\phi_{g}\left({\bf r}_{a},t_{1}\right).\nonumber\\
\label{2nd_pulse}
\end{eqnarray}
The exponential describing the effect of the second pulse can
also be expanded in a truncated series of Bessel functions and 
we have.
\begin{eqnarray}
\phi_{g}\left({\bf r}_{a},t_{2+}\right) &\approx&
e^{i\Omega_{2}(\delta t_{1}+\delta t_{2})}\nonumber\\
&\times&
\sum_{n=-n_{max}}^{n_{max}}\,
\sum_{n^{\prime}=-n_{max}}^{n_{max}}\,
i^{n+n^{\prime}}
J_{n}\left(\Omega_{2}\delta t_{1}\right)\nonumber\\
&\times&
J_{n^{\prime}}\left(\Omega_{2}\delta t_{2}\right)
e^{-iE_{n}(t_{2}-t_{1})/\hbar}
e^{2\left(n+n^{\prime}\right)i{\bf k}_{L}\cdot{\bf r}_{a}}\nonumber\\
&\times&
\phi_{g}\left({\bf r}_{a},t_{1}\right).
\end{eqnarray}
Changing summation indexes as $m\equiv n + n^{\prime}$ we can write
\begin{eqnarray}
\phi_{g}\left({\bf r}_{a},t_{2+}\right) &\approx&
e^{i\Omega_{2}(\delta t_{1}+\delta t_{2})}\nonumber\\
&\times&
\sum_{m=-2n_{max}}^{2n_{max}}\,
A_{m}e^{2mi{\bf k}_{L}\cdot{\bf r}_{a}}
\phi_{g}\left({\bf r}_{a},t_{1}\right),\nonumber\\
\end{eqnarray}
where
\begin{equation}
A_{m} \equiv i^{m}
\sum_{n=-\infty}^{\infty}
J_{m-n}\left(\Omega_{2}\delta t_{2}\right)
J_{n}\left(\Omega_{2}\delta t_{1}\right)
e^{-iE_{n}(t_{2}-t_{1})/\hbar}.
\label{two-pulse_amplitude}
\end{equation}
Note that we have extended the limits of the summation back to infinity.
This is possible because, in the Raman--Nath regime, the values of 
$J_{n}\left(\Omega_{2}\delta t\right)$ for $n > n_{max}$ are sufficiently 
small that we incur little error in including these extra terms.  So, 
finally, we have
\begin{eqnarray}
\phi_{g}\left({\bf r}_{a},t\right) &\approx&
e^{i\Omega_{2}(\delta t_{2}+\delta t_{1})}\nonumber\\
&\times&
\sum_{m=-\infty}^{\infty}\,
A_{m}e^{2mi{\bf k}_{L}\cdot{\bf r}_{a}}
\phi_{g}\left({\bf r}_{a},t_{1}\right),\nonumber\\
\end{eqnarray}
The quantity $A_{m}$ is the probability amplitude for an atom
to be in momentum state ${\bf p}_{m} = 2m\hbar{\bf k}_{L}$.

\begin{figure*}[htb]
\begin{center}
\mbox{\psboxto(6.0in;0in){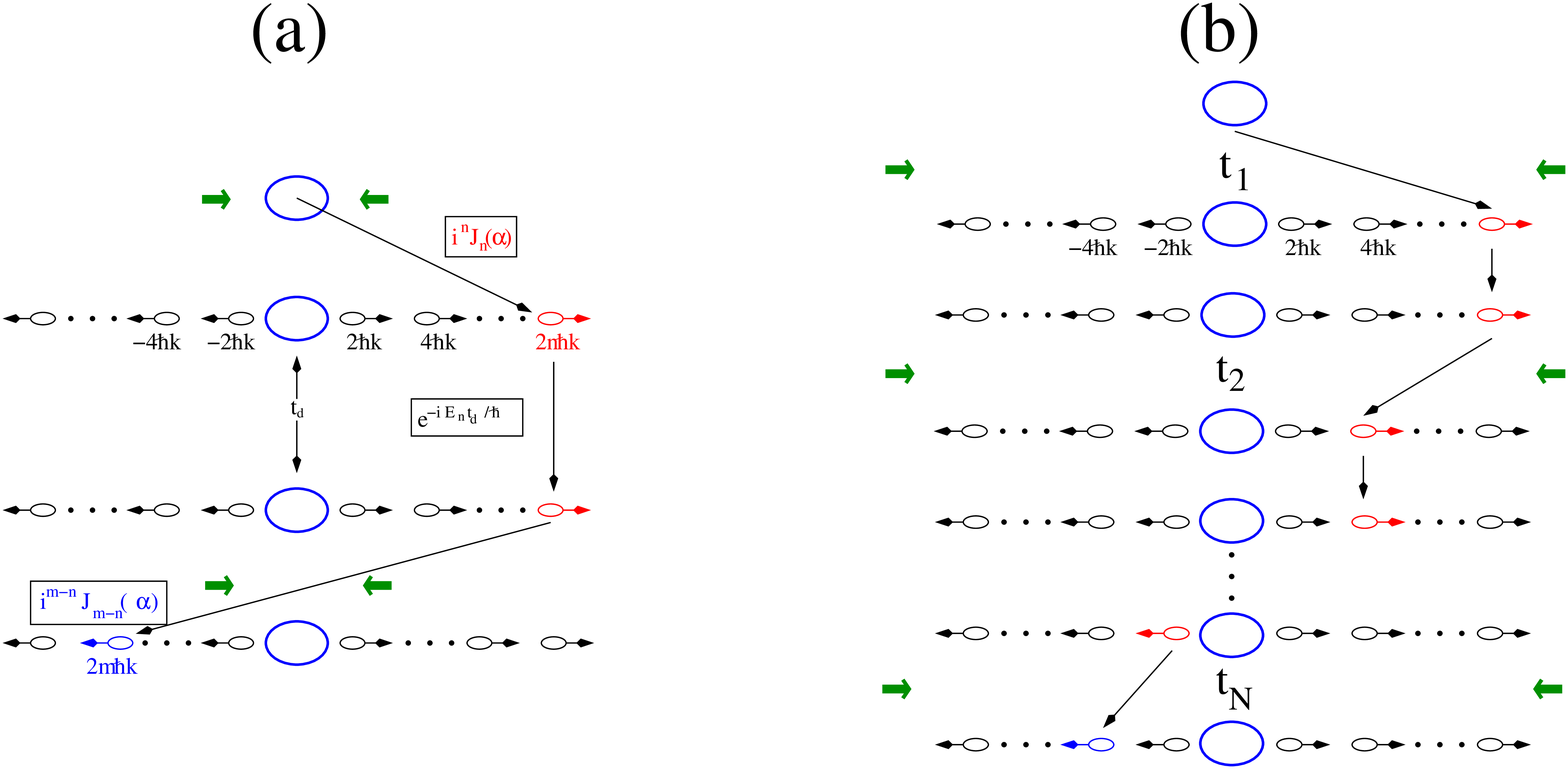}}
\end{center}
\caption{(a) (color online) The final distribution of the condensate 
atoms over the momentum states after the application of two standing--wave
pulses separated by an interval can be understood in terms of 
interfering quantum pathways.  The amplitude for the system to
jump from the initial zero--momentum state at the top to the final 
state at the bottom along the path shown is the product of the 
amplitudes (shown in boxes) for the three legs that compose the 
path. The total amplitude is coherent summation of this composite
single--path amplitude over all possible paths. (b) The general case 
of $N$ pulses is illustrated in the right panel. We assume that 
pulse 1 is applied at $t_{1}$, pulse 2 at $t_{2}$, etc.}
\label{interp_fig}
\end{figure*}

\section{General features of the solution for two pulses}
\label{general}

Before deriving some general features of the two--pulse amplitude,
it will be convenient to rewrite Eq.\ (\ref{two-pulse_amplitude}) in
a form where the time interval is measured in appropriate units. 
To this end we define the period, $T_{T}$, of free oscillation of 
a free particle whose momentum is ${\bf p}_{1} = 2\hbar{\bf k}_{L}$ 
such that:
\begin{equation}
\left(\frac{p_{1}^{2}}{2m}\right)T_{T}/\hbar = 2\pi.
\end{equation}
This is the Talbot time defined above. In this case
\begin{eqnarray}
\frac{E_{n}(t_{2}-t_{1})}{\hbar} &=& 
n^{2}\left(\frac{p_{1}^{2}}{2m}\right)
\frac{(t_{2}-t_{1})}{\hbar}\nonumber\\
&=&
\frac{2\pi n^{2}(t_{2}-t_{1})}{T_{T}} 
\equiv
2\pi\beta_{1}n^{2},
\end{eqnarray}
where we have introduced the {\em time interval} parameter as 
$\beta_{1}\equiv (t_{2}-t_{1})/T_{T}$, that is, the time between pulses
measured in units of the Talbot time.  Also defining the {\em pulse area} 
parameters as $\alpha_{1}\equiv\Omega_{2}\delta t_{1}$ and 
$\alpha_{2}\equiv\Omega_{2}\delta t_{2}$, we can write $A_{m}$ as
\begin{equation}
A_{m}\left(\alpha_{1},\alpha_{2},\beta_{1}\right) = i^{m}
\sum_{n=-\infty}^{\infty}
J_{m-n}\left(\alpha_{2}\right)
e^{-2\pi i n^{2}\beta_{1}}
J_{n}\left(\alpha_{1}\right).
\label{amp}
\end{equation}
We will use this form of the two--pulse amplitude to demonstrate
some its general features.

\begin{figure*}[htb]
\begin{center}
\mbox{\psboxto(\textwidth;0in){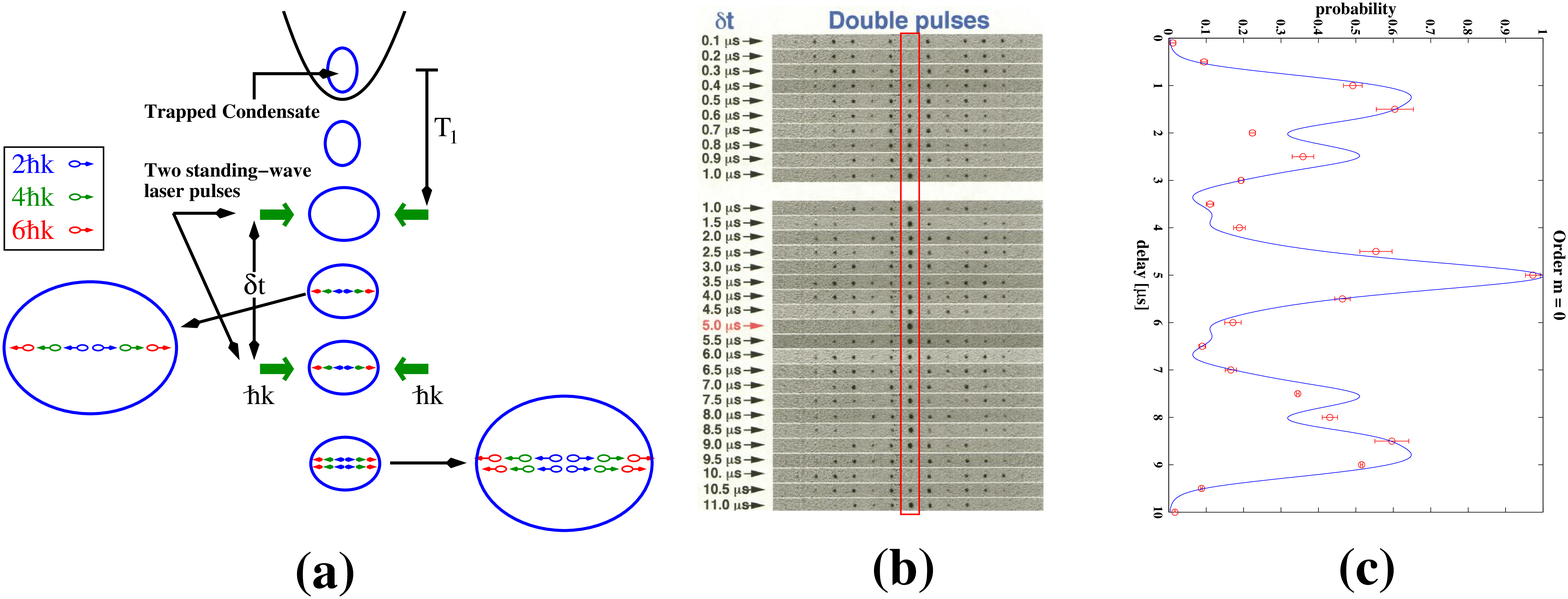}}
\end{center}
\caption{(color online) Comparison of the theory presented in the 
text with the results of the temporal Talbot experiment reported in 
Ref.\ \cite{PhysRevLett.83.5407}. (a) The left panel shows the
steps of this experiment.  Two 100--ns, 1 W/cm$^{2}$,
standing--wave laser pulses were applied to a $^{23}$Na BEC after 
allowing the condensate to expand for 1.2 ms.  The pulses were 
separated in time by an interval, $\delta t$, which varied between 0.1 and
11 $\mu$s. (b) The middle panel shows the images taken after 
allowing the condensate to expand for a further 6.2 ms.  Each column
of dots corresponds to a specific final momentum state.  The highlighted
column corresponds to zero momentum while the column just to the 
right (left) of it corresponds to the $2\hbar k$ ($-2\hbar k$) momentum
state. (c) The right panel shows a comparison of the prediction of 
Eq.\ (\ref{two-pulse_amplitude}) with the normalized pixel count of 
the dots of the highlighted column.}
\label{exp_cmp}
\end{figure*} 

\subsection{Symmetry of the momentum distribution}
\label{symmetry}

There are two general symmetries exhibited by the expression
in Eq.\ (\ref{amp}).  The first symmetry is that the amplitudes 
for opposite momentum orders are equal for a given pulse 
sequence, that is 
\begin{equation}
A_{m}(\alpha_{1},\alpha_{2},\beta_{1}) = 
A_{-m}(\alpha_{1},\alpha_{2},\beta_{1}).
\label{m_sym}
\end{equation}

To show this symmetry, we write the expression for 
$A_{-m}(\alpha_{1},\alpha_{2},\beta_{1})$ which is
\begin{eqnarray}
A_{-m}\left(\alpha_{1},\alpha_{2},\beta_{1}\right) 
&=& i^{-m}
\sum_{n=-\infty}^{\infty}
J_{-m-n}\left(\alpha_{2}\right)\nonumber\\
&\times&
e^{-2\pi i n^{2}\beta_{1}}
J_{n}\left(\alpha_{1}\right).
\label{a_minus_m}
\end{eqnarray}
Using the Bessel generating function~\cite{NIST_DLMF}, it is easy to show
that $J_{-n}(z) = (-)^{n}J_{n}(z)$ so that
\begin{eqnarray*}
A_{-m}\left(\alpha_{1},\alpha_{2},\beta_{1}\right) 
&=& i^{-m}
\sum_{n=-\infty}^{\infty}
(-)^{m+n}J_{m+n}\left(\alpha_{2}\right)\\
&\times&
e^{-2\pi i n^{2}\beta_{1}}
(-)^{n}J_{-n}\left(\alpha_{1}\right)\\
&=& i^{m}
\sum_{n=-\infty}^{\infty}
J_{m+n}\left(\alpha_{2}\right)\\
&\times&
e^{-2\pi i n^{2}\beta_{1}}
J_{-n}\left(\alpha_{1}\right).
\end{eqnarray*}
Changing the summation index to $n^{\prime}=-n$ we have
\begin{eqnarray*}
A_{-m}\left(\alpha_{1},\alpha_{2},\beta_{1}\right) 
&=& i^{m}
\sum_{n^{\prime}=-\infty}^{\infty}
J_{m-n^{\prime}}\left(\alpha_{2}\right)\\
&\times&
e^{-2\pi i (-n^{\prime})^{2}\beta_{1}}
J_{n^{\prime}}\left(\alpha_{1}\right),
\end{eqnarray*}
and so
\begin{equation}
A_{-m}\left(\alpha_{1},\alpha_{2},\beta_{1}\right) = 
A_{m}\left(\alpha_{1},\alpha_{2},\beta_{1}\right).
\label{m_symmetry}
\end{equation}
Thus all distributions are exactly symmetric with respect to momentum
order under these conditions.  Physically when atoms
make a transition to a non--zero momentum state they must absorb photons
from one laser beam and emit into the other beam. Equation (\ref{m_symmetry})
holds because, for standing--wave laser beams, there is no preference
for choosing which beam photons are absorbed and into which emitted.
Absorbing from opposite beams results in populating opposite final momentum 
states.

The second symmetry is that the amplitude for a given pair of pulses 
separated by an interval $\beta_{1}$ is the complex conjugate of the
amplitude for an interval of $1-\beta_{1}$ for fixed $\alpha_{1}$ and
$\alpha_{2}$.  Thus we have
\begin{equation}
A_{m}(\alpha_{1},\alpha_{2},1-\beta_{1}) = 
A_{m}^{\ast}(\alpha_{1},\alpha_{2},\beta_{1})
\qquad
\beta_{1} \le 1/2.
\label{beta_sym}
\end{equation}
If we only consider interval times $\beta_{1} \le 1/2$, it is easy to
see that this holds in Eq.\ (\ref{two-pulse_amplitude}).  Since
$e^{-2\pi i n^{2}(1-\beta_{1})} = e^{2\pi i n^{2}\beta_{1}}$ gives the
complex conjugate of the exponential in the $\beta_{1}$ amplitude.
Given the fact that the Bessel functions are all real, conjugating the
exponential conjugates the entire expression.  Equation (\ref{beta_sym})
also implies that the probability distribution for two--pulse sequences
with intervals $\beta_{1}$ and $1-\beta_{1}$ are identical.  This result
will be useful later to constrain the parameter space in the momentum
space design procedure described below.

\subsection{Special values of the time interval}
\label{special_vals}

Two important results can be obtained for the quantity 
$A_{m}(\alpha_{1},\alpha_{2},\beta_{1})$ when the interval time has 
special values. The two special cases are (1) $\beta_{1} = \frac{1}{2}$, 
and (2) $\beta_{1} = 1$.  We consider each of these in turn.

The first case occurs when the time interval between standing--wave 
pulses equals half of the Talbot time.  Here, the exponential factor in 
Eq.\ (\ref{amp}) becomes
\begin{equation}
e^{-2\pi i n^{2}\beta_{1}} = 
e^{-\pi i n^{2}} = 
\left(-\right)^{n^{2}} = 
\left(-\right)^{n}.
\end{equation}
The last equality can be seen by noting that the square of an 
even integer is even and the square of an odd integer is odd.
Thus, the probability amplitude becomes
\begin{equation}
A_{m}\left(\alpha_{1},\alpha_{2},\frac{1}{2}\right) = i^{m}
\sum_{n=-\infty}^{\infty}
\left(-\right)^{n}
J_{m-n}\left(\alpha_{2}\right)
J_{n}\left(\alpha_{1}\right).
\label{amp_half}
\end{equation}
This expression can be summed exactly as follows.  We write
again the Bessel generating function
\begin{eqnarray}
e^{\frac{1}{2}\alpha_{1}\left(t - \frac{1}{t}\right)} &=&
\sum_{n=-\infty}^{\infty}\,
t^{n}J_{n}\left(\alpha_{1}\right)\nonumber\\
e^{-\frac{1}{2}\alpha_{2}\left(t - \frac{1}{t}\right)} &=&
\sum_{n^{\prime}=-\infty}^{\infty}\,
\left(-\right)^{n^{\prime}}t^{n^{\prime}}
J_{n^{\prime}}\left(\alpha_{2}\right).
\end{eqnarray}
In the second equality, we have let $\alpha_{1} \rightarrow -\alpha_{2}$ 
and used the identity $J_{n}\left(-\alpha\right) = \left(-\right)^{n}
J_{n}\left(\alpha\right)$.  Equating the product of the right--hand--sides 
with the product of the left--hand--side of the above two equations and
letting $m = n + n^{\prime}$ yields the following
\begin{eqnarray}
e^{-\frac{1}{2}\left(\alpha_{2}-\alpha_{1}\right)
\left(t-\frac{1}{t}\right)} 
&=& 
\sum_{m=-\infty}^{\infty}i^{m}
A_{m}\left(\alpha_{1},\alpha_{2},\frac{1}{2}\right)t^{m}\nonumber\\
&=&
\sum_{m=-\infty}^{\infty}\left(-\right)^{m}
J_{m}\left(\alpha_{2}-\alpha_{1}\right)t^{m},\nonumber\\
\end{eqnarray}
where the second equality comes from a direct application of the Bessel 
generating function to the exponential on the left--hand--side. Since the 
equality of the two sums must hold for any value of $t$, the coefficients 
of $t^{m}$ in the sums must be equal.  And so,
\begin{equation}
A_{m}\left(\alpha_{1},\alpha_{2},\frac{1}{2}\right) = 
i^{m}J_{m}\left(\alpha_{2}-\alpha_{1}\right)
\label{amp_one_half}
\end{equation}
One consequence of this is that, for equal--area pulses and when the 
interval between pulses is one--half of the Talbot time, (that is,
when $\alpha_{1}=\alpha_{2}$ and $\beta_{1}=1/2$) the amplitudes for
all non--zero momentum states are zero and the condensate is unchanged.
This effect was verified experimentally and reported in 
Ref.\ \cite{PhysRevLett.83.5407}.  Figure \ref{exp_cmp}(b) shows this
effect where a double--pulse delay of 5 $\mu$s results in no change to 
the original condensate.

In the second case where $\beta_{1} = 1$ the exponential in the expression 
for $A_{m}(\alpha_{1},\alpha_{2},1)$ equals unity for all values of the 
summation index $n$.  Thus we can write
\begin{eqnarray}
A_{m}\left(\alpha_{1},\alpha_{2},1\right) 
&=& i^{m}
\sum_{n=-\infty}^{\infty}
J_{m-n}\left(\alpha_{2}\right)
J_{n}\left(\alpha_{1}\right)\nonumber\\
&=&
i^{m}J_{m}\left(\alpha_{2}+\alpha_{1}\right)
\label{amp_one}
\end{eqnarray}
where the second equality is derived by a method similar to that which
produced Eq.\ (\ref{amp_one_half}).  This result suggests that two  
standing--wave pulses separated in time by one Talbot time have the 
same effect as a single pulse whose area is the sum of the areas of 
the two pulses.

\subsection{Physical interpretation of the probability amplitude}
\label{interp}

The general formula for the probability amplitude, Eq.\ %
(\ref{two-pulse_amplitude}), can be understood as the superposition 
of amplitudes of multiple pathways from the given initial to the 
given final state. One such pathway is illustrated in Fig.\ %
\ref{interp_fig}(a). This figure depicts a particular quantum pathway 
from the fixed initial zero--momentum state to a fixed final state 
whose momentum is $p_{m}=2m\hbar k$. The amplitude for an atom to go
between these states via the path shown is the product of the amplitudes 
for the three legs of the path.  The first leg is a momentum jump, 
caused by the first pulse, from the zero--momentum state to the momentum 
state $p = p_{n} = 2n\hbar k$ with amplitude $i^{n}J_{n}\left(\alpha\right)$.  
In the second leg of the path, atoms in the momentum state $p=p_{n}$, 
whose energy is $E_{n}=p_{n}^{2}/2m$, evolve as free particles during the 
short--time interval between pulses. Thus the amplitude to ``jump'' from 
the time just after the first pulse to just before the second pulse is 
$e^{-iE_{n}t_{d}/\hbar}$.  The final leg of the path shown is another 
momentum jump, caused by the second pulse, from $p = p_{n}$ to $p = p_{m}$ 
and whose amplitude is $i^{m-n}J_{m-n}(\alpha)$.  The amplitude 
to proceed from the initial to the final state is the product of the
amplitudes for the three legs.

This pathway proceeds from the initial state to the final state
via the momentum state $p_{n}$.  The system can make the transition
between these initial and final states via any state $p_{n}$ and
since these different pathways are not detected, the total amplitude
for the system to jump from the initial to the final state is the 
coherent summation of these individual amplitudes given in 
Eq.\ (\ref{two-pulse_amplitude}).  This quantum pathways interpretation
will enable an easy generalization of the two--pulse amplitude to
the $N$--pulse case.

\subsection{Comparison with experiment}
\label{exp_compare}

The validity of this theory for two pulses was tested in an experiment
and reported in Ref.\ (\cite{PhysRevLett.83.5407}).  In this experiment,
a BEC consisting of $3\times 10^{6}$ Na atoms confined in the 
$F=1,m_{F}=-1$ ground state by a time--averaged orbiting potential
(TOP) trap~\cite{PhysRevLett.74.3352} were released and allowed to 
expand for 1.2 ms as illustrated in the left panel of Fig.\ (\ref{exp_cmp}).
Next, two 589--nm--wavelength, 100--ns--duration, standing--wave, 
linearly polarized laser pulses were applied with a time interval between them 
which varied between 1 and 10 $\mu$s.  The intensity of the pulses was 
about 1 W/cm$^{2}$ and they were detuned by approximately 600 MHz from
the $3S_{1/2},F=1\rightarrow 3P_{3/2},F^{\prime}=2$ transition. The 
condensate was then allowed to expand for a further 6.2 ms at which time 
an absorption image was taken.  This last expansion enabled atoms in 
non--zero momentum states to leave the condensate and the resulting image 
is a measurement of the momentum--space distribution immediately after 
the second laser pulse. 

The middle panel in Fig.\ (\ref{exp_cmp}) shows the results of these
absorption images for varying interval times between the pulses.  Each row
shows a picture of the data for a particular interval time while the columns
indicate particular momentum states.  The center column highlighted
shows atoms in the zero--momentum state.  The graph shown on the right--hand
panel is a comparison of the normalized pixel counts of the dots in
the zero--momentum states (highlighted column in panel (b)) with the 
theory curve $|A_{0}(\alpha_{0},\alpha_{0},\beta)|^{2}$ plotted as a function
of $\beta$ and where $\alpha_{0}$ is the product of the two--photon
Rabi frequency and the pulse time corresponding to the experimental
conditions.  It is important to note that there are no adjustable 
parameters in this calculation.  One can see that there is good agreement
between theory and experiment which in turn lends support for the model
presented above. 

\begin{figure*}[htb]
\begin{center}
\mbox{\psboxto(\textwidth;0in){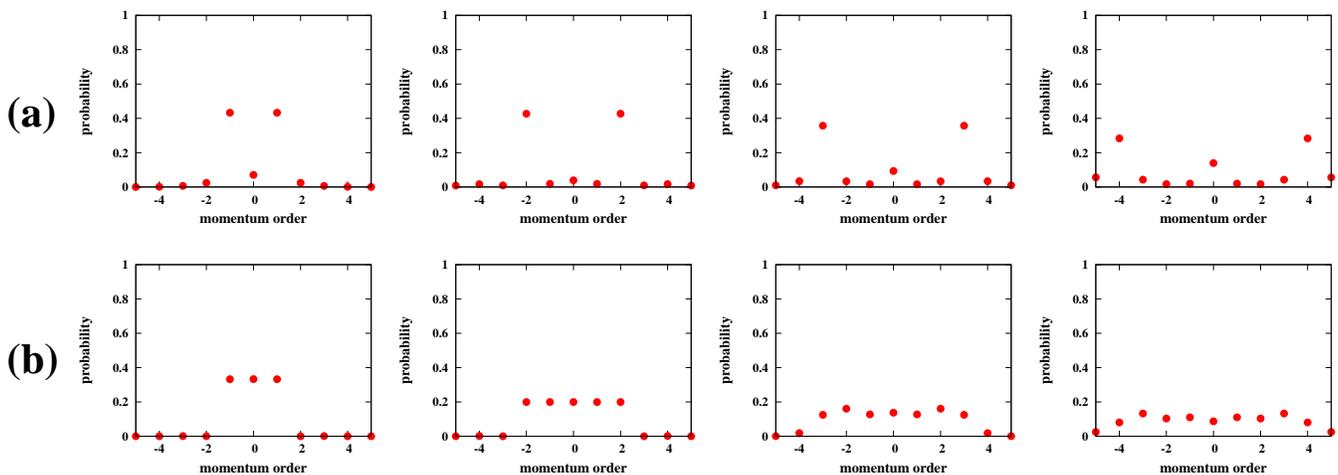}}
\end{center}
\caption{(color online) The figure above shows examples of shaping 
momentum distributions using only two pulses.  The shaping parameters 
are the pulse areas, $\alpha_{1}$ and $\alpha_{2}$, and the interval 
between them, $\beta_{1}$.  All of the plots above show the 
probabilities of atoms being in momentum orders $2m\hbar k$ where 
$-5 \le m \le 5$. Each plot is annotated with the values of 
$\alpha_{1}$, $\beta_{1}$, and $\alpha_{2}$ that produced each 
momentum distribution. (a) The four plots in the top panel exhibit 
(from left to right) the distributions $D_{mag}^{(2)}(m)$ where 
the two momentum states $-2m\hbar k$ and $+2m\hbar k$ are equally 
populated and where $1 \le m \le 4$. (b) The four plots in the 
bottom panel depict (again from left to right) distributions 
$D_{range}^{(2)}(m)$ where all of the states in the range between 
$-2m\hbar k$ and $+2m\hbar k$ are equally populated and where 
$1 \le m \le 4$.}
\label{shaping_fig1}
\end{figure*} 

\section{$N$--pulse theory}
\label{n-pulse_theory}

Consider a condensate that is subjected to a sequence of $N$ pulses with
arbitrary pulse areas and times.  As shown in Fig.\ \ref{interp_fig}(b), 
we assume that pulse 1 is applied at $t=t_{1}$ and has pulse area 
$\alpha_{1}$, pulse 2 at $t=t_{2}$ with area $\alpha_{2}$, $\dots\ $, 
pulse $N$ at $t=t_{N}$ with area $\alpha_{N}$.  If we label the momentum 
state jumped to at the time of pulse $k$ as $n_{k}$, where $0 \le k \le N$,
then a single quantum pathway through the entire $N$--pulse sequence can
be labeled by specifying the index of the momentum state the system jumps
to after each pulse: $(n_{1},n_{2},\dots,n_{N-1})$.  Then it is easy to 
write the amplitude for a particular $N$--pulse pathway by analogy with 
the two--pulse case.  The resulting amplitude for a single quantum pathway 
whose starting momentum state is $2n_{0}\hbar k$ and whose ending state is 
$2n_{N}\hbar k$ for an $N$--pulse sequence labeled in this way is
\begin{eqnarray}
A_{n_{N},n_{0}}^{(N)}({\mb\alpha},{\mb \delta t},{\mb n})
&=&
(i^{n_{1}-n_{0}}J_{n_{1}-n_{0}}
(\alpha_{1})e^{-iE_{n_{1}}(t_{2}-t_{1})/\hbar})\nonumber\\
&\times&
(i^{n_{2}-n_{1}}J_{n_{2}-n_{1}}
(\alpha_{2})e^{-iE_{n_{2}}(t_{3}-t_{2})/\hbar})\nonumber\\
&\times&\cdots
(i^{n_{N}-n_{N-1}}J_{n_{N}-n_{N-1}}
(\alpha_{N})),
\label{n-pulse_path}
\end{eqnarray}
where we have assumed that the initial momentum state is not zero but
rather labeled by $n_{0}$ and we have not considered any interval
following the final pulse at $t=t_{N}$.  

The vectors ${\mb\alpha}$ and ${\mb \delta t}$ label the areas and 
intervals of the applied pulse sequence.
\begin{eqnarray}
{\mb\alpha} &\equiv& 
\left(\alpha_{1},\alpha_{2},\dots,\alpha_{N}\right)\\
{\mb\delta t} &\equiv&
\left(\delta t_{1}=t_{2}-t_{1},\dots,\delta t_{N-1}=t_{N}-t_{N-1}\right).
\end{eqnarray}
It will be convenient to rescale the intervals in units of the Talbot
time as was done for two pulses.  Thus we define $\beta_{k}=\delta_{k}/T_{T}$
for $1 \le k \le N-1$.  Hence we can rewrite Eq.\ (\ref{n-pulse_path})
as
\begin{eqnarray}
A_{n_{N},n_{0}}^{(n_{1},\dots,n_{N-1})}({\mb\alpha},{\mb\beta}) 
&=&
i^{n_{N}-n_{0}}
J_{n_{1}-n_{0}}(\alpha_{1})
e^{-2\pi i n_{1}^{2}\beta_{1}}\nonumber\\
&\times&
J_{n_{2}-n_{1}}(\alpha_{2})
e^{-2\pi i n_{2}^{2}\beta_{2}}\cdots\nonumber\\
&\times&
e^{-2\pi i n_{N-1}^{2}\beta_{N-1}}
J_{n_{N}-n_{N-1}}(\alpha_{N}),\nonumber\\
\label{n-pulse_path_2}
\end{eqnarray}

To get the full amplitude to jump from the initial momentum state, labeled
by $n_{0}$, to the final momentum state, labeled by $n_{N}$, we coherently
sum over all of the single--path amplitudes.  This yields the following.
\begin{eqnarray}
A_{n_{N},n_{0}}^{(N)}({\mb\alpha},{\mb\beta}) 
&=&
\sum_{n_{1}}\cdots\sum_{n_{N-1}}
A_{n_{N},n_{0}}^{(n_{1},\dots,n_{N-1})}({\mb\alpha},{\mb\beta})\nonumber\\
&=&
i^{n_{N}-n_{0}}\sum_{n_{1}}\cdots\sum_{n_{N-1}}
J_{n_{1}-n_{0}}(\alpha_{1})\nonumber\\
&\times&
e^{-2\pi i n_{1}^{2}\beta_{1}}
J_{n_{2}-n_{1}}(\alpha_{2})
e^{-2\pi i n_{2}^{2}\beta_{2}}\nonumber\\
&\cdots&
e^{-2\pi i n_{N-1}^{2}\beta_{N-1}}J_{n_{N}-n_{N-1}}(\alpha_{N}).
\label{full_amplitude}
\end{eqnarray}
This is the general result for the amplitude to jump from an initial
momentum state of $p_{i}=2n_{0}\hbar k$ to a final momentum state of
$p_{f}=2n_{N}\hbar k$ due to the application of $N$ pulses whose areas
are $\alpha_{1},\dots,\alpha_{N}$ separated by $N-1$ intervals of 
durations (expressed in Talbot--time units) $\beta_{1},\dots,\beta_{N-1}$.
It holds as long as the {\em Raman--Nath} approximation is valid for
all pulses, i.e., that atoms in non--zero--order momentum states do not
move an appreciable distance compared to the condensate size during the
entire pulse sequence.  We can now use this result to design sequences 
of such pulses and intervals to engineer a specified momentum--state 
probability distribution.

\section{Engineering momentum--state probability distributions}
\label{eng_dist}

\subsection{Least--squares design}
\label{least_squares}

Designing a pulse sequence to engineer a specified probability 
distribution across the momentum states $2m\hbar k$ is straightforward.
First, the momentum distribution is described by specifying the 
set of desired probabilities $\{p_{m}\}$ for momentum orders $2m\hbar k$ 
for all $m$.  We will refer to this set of numbers as the {\em momentum 
probability distribution}.  This set of numbers must satisfy several 
conditions to be a valid momentum probability distribution.  Each $p_{m}$ 
must be a probability and the distribution must be normalized so that
\begin{equation}
0 \le p_{m} \le 1,
\qquad
-\infty < m < \infty,
\end{equation}
and
\begin{equation}
\sum_{m=-\infty}^{\infty}p_{m} = 1.
\end{equation}
Furthermore, since all of the pulses are assumed to be standing waves, 
by symmetry the probability for $2m\hbar k$ must equal that for 
$-2m\hbar k$ and thus
\begin{equation}
p_{m} = p_{-m},
\qquad
-\infty < m < \infty.
\end{equation}

Once the momentum probability distribution is specified, the least
squares procedure can be carried out by defining the $N$--pulse,
least--squares cost function:
\begin{equation}
F_{LS}^{(N)}({\mb\alpha},{\mb\beta}) \equiv 
\sum_{m=-\infty}^{\infty}
\left[
p_{m}-\left|A_{m,0}^{(N)}({\mb\alpha},{\mb\beta})\right|^{2}
\right]^{2}
\end{equation}
and finding the values of the parameters ${\mb\alpha}={\mb\alpha}^{min}$
and ${\mb\beta}={\mb\beta}^{min}$ that produces the global minimum of
$F_{LS}({\mb\alpha},{\mb\beta})$.  

\begin{figure*}[htb]
\begin{center}
\mbox{\psboxto(\textwidth;0in){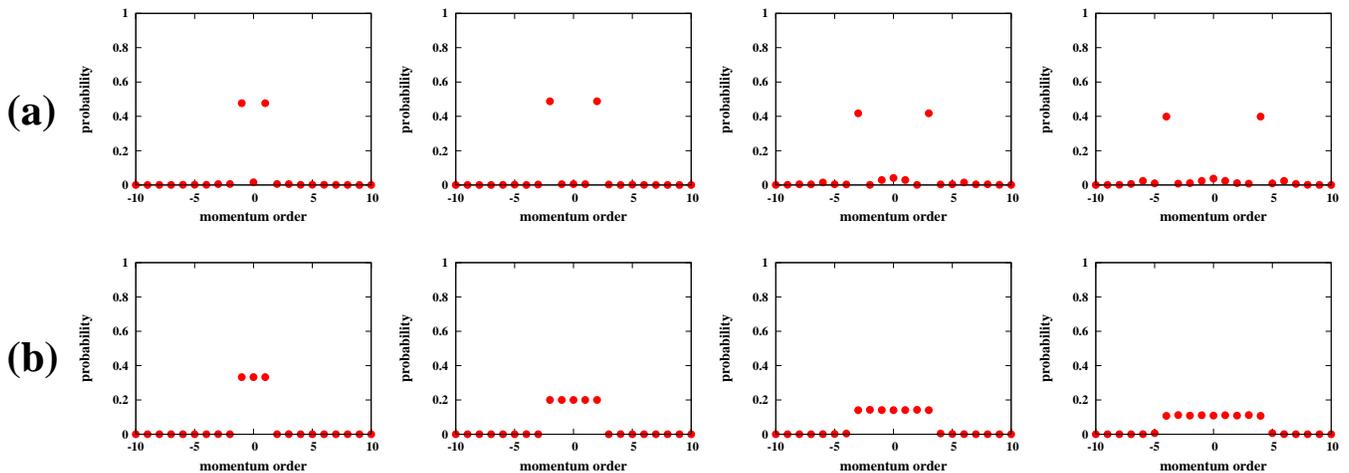}}
\end{center}
\caption{(color online) These plots show examples of momentum distributions 
shaped by three pulses.  The examples shown are the same as those shown in
Fig.\ \ref{shaping_fig1} for two pulses.  The shaping parameters are 
the areas of the three pulses, $\alpha_{1}$, $\alpha_{2}$, and $\alpha_{3}$, 
and the intervals between them, $\beta_{1}$ and $\beta_{2}$.  All of 
the plots above show the probabilities of atoms being in momentum orders 
$2m\hbar k$ where $-10 \le m \le 10$. (a) The four plots in the top 
panel exhibit (from left to right) the distributions $D_{mag}^{(3)}(m)$ 
where the two momentum states $-2m\hbar k$ and $+2m\hbar k$ are equally 
populated and where $1 \le m \le 4$. (b) The four plots in the 
bottom panel depict (again from left to right) distributions 
$D_{range}^{(3)}(m)$ where all of the states in the range between 
$-2m\hbar k$ and $+2m\hbar k$ are equally populated and where 
$1 \le m \le 4$.}
\label{shaping_fig2}
\end{figure*} 

In general, this minimization must be done numerically. It is also subject 
to certain constraints which both derive from the Raman--Nath approximation.
First, we assume that all intervals are less than or equal to one Talbot 
time, or $0\le\beta_{n}\le 1$ for $1 \le n \le N$.  Second, there is a limit 
on the size of each individual pulse area because too large an area will 
produce population in such a high momentum state that there will be 
appreciable motion during the pulse sequence.  Thus there is a maximum 
value, $\alpha_{max}$, such that $0\le\alpha_{n}\le\alpha_{max}$.  These
constraints also limit the total number of pulses that can be practically
applied while still satisfying all of the conditions described above. 
However, as we shall see, many important momentum probability distributions
can be achieved with high fidelity using only two or three pulses.  

We implemented the least--squares minimization procedure described above 
to design two--pulse and three--pulse sequences to produce momentum 
distributions in two categories. The first category is a distribution where 
all of the population appears in the two momentum states where the magnitude 
is $2n\hbar k$ (that is, $+2n\hbar k$ or $-2n\hbar k$) or equivalently 
where $p_{n}=p_{-n}=1/2$.  We will use the label, $D_{mag}^{(N)}(n)$, to 
stand for the distribution determined from the $N$--pulse, least--squares 
procedure when the desired distribution is of this type.  The other kind 
of distribution we considered has equal populations in the range of momentum 
states between $\pm 2n\hbar k$. In this case $p_{-n}=\dots=p_{n}=1/(2n+1)$. 
The distribution produced by the $N$--pulse, least--squares procedure when 
this type of distribution will be labeled by $D_{range}^{(N)}(n)$.  

These categories of distributions are interesting because they act like
two-- and multiple--beam splitters.  The ability to transfer condensate
population into these kinds of momentum distributions might be useful
in designing atom interferometry experiments or for quantum information
processing.  In this regard we are particularly interested in discovering 
how high a ``fidelity'' can be achieved with just a few pulses. 

Hereafter we will take the measure of the ``fidelity,'' that is, how 
close the actual momentum probability distribution is to the specified 
one,  to be the minimum value of the least--squares cost function. Thus,
\begin{equation}
\left(F_{LS}^{(N)}\right)_{min} \equiv
F_{LS}^{(N)}\left({\mb\alpha}^{min},{\mb\beta}^{min}\right)
\end{equation}
will be used to measure the fidelity.

\begin{table*}
\caption{The table below presents the optimal values of the parameters and
the values of the resulting momentum distribution probabilities for two-- 
and three--pulse sequences for the momentum magnitude and momentum range 
distributions contained in Figs.\ \ref{shaping_fig1} and \ref{shaping_fig2}.  
These values were determined by the least--squares procedure discussed in 
the text. The leftmost column lists the specified momentum distributions in
boldface.  The next five columns exhibit the parameter values for two--pulse 
sequences while the final seven columns give the three--pulse--sequence 
results.  The two--pulse parameters are the pulse areas, $\alpha_{1}$ and 
$\alpha_{2}$, and the interval, $\beta_{1}$, expressed in units of the Talbot 
time.  The three--pulse parameters are the areas of the three pulses, 
$\alpha_{1}$, $\alpha_{2}$, and $\alpha_{3}$, and the two intervals, $\beta_{1}$ 
and $\beta_{2}$.  Columns two and seven, labeled by $p_{n}^{(opt)}$, give 
the optimal--distribution values of the nonzero specified probabilities. The
probability given is listed in the first column in lightface type.
Columns six and thirteen list the fidelity of the least--squares momentum 
distributions to the specified distribution by giving the value of the 
least--squares functions $F_{LS}^{(2)}$ and $F_{LS}^{(3)}$ evaluated at the 
values of the parameters listed in the table.}
\begin{tabular}{||c||ccccc||ccccccc||}
\hline
\multicolumn{1}{||c||}{Momentum}&
\multicolumn{5}{c||}{Two--pulse results}&
\multicolumn{7}{c||}{Three--pulse results}\\
Distribution &
\ \ \ $p_{n}^{(opt)}$\ \ \  &\ \ \ $\alpha_{1}^{min}$\ \ \ &
\ \ \ $\beta_{1}^{min}$\ \ \ &\ \ \ $\alpha_{2}^{min}$\ \ \ &
$(F_{LS}^{(2)})_{min}$
&$p_{n}^{(opt)}$&\ $\alpha_{1}^{min}$\ &
\ $\beta_{1}^{min}$\ &
\ $\alpha_{2}^{min}$\ &
\ $\beta_{2}^{min}$\ &
\ $\alpha_{3}^{min}$\ &
$(F_{LS}^{(3)})_{min}$\\
\hline
${\mb p}_{-1}={\mb p}_{1}=\frac{1}{2}$      &        &1.715 & 0.130 & 0.594 & $1.5\times10^{-2}$
&        & 3.737 & 0.481 & 3.402 & 0.453 & 1.782 & $1.9\times10^{-3}$\\
$p_{-1}^{(opt)}=p_{1}^{(opt)}$  & 0.4327 &&&&& 0.4766 &&&&&&\\
${\mb p}_{-2}={\mb p}_{2}=\frac{1}{2}      $&        &2.857 & 0.320 & 1.04 & $1.4\times10^{-2}$
&        & 2.733 & 0.342 & 0.585 & 0.337 & 0.882 & $3.9\times10^{-4}$\\
$p_{-2}^{(opt)}=p_{2}^{(opt)}$  & 0.4271 &&&&& 0.4877 &&&&&&\\
${\mb p}_{-3}={\mb p}_{3}=\frac{1}{2}      $&        &3.560 & 0.378 & 1.429 & $5.6\times10^{-2}$
&        & 1.297 & 0.586 & 7.347 & 0.404 & 2.310 & $1.7\times10^{-2}$\\
$p_{-3}^{(opt)}=p_{3}^{(opt)}$  & 0.3572 &&&&& 0.4179 &&&&&&\\
${\mb p}_{-4}={\mb p}_{4}=\frac{1}{2}      $&        &4.230 & 0.408 & 1.790 & $1.2\times10^{-1}$
&        & 3.158 & 0.376 & 2.936 & 0.287 & 1.047 & $2.5\times10^{-2}$\\
$p_{-4}^{(opt)}=p_{4}^{(opt)}$  & 0.2833 &&&&& 0.3984 &&&&&&\\
${\mb p}_{-1}={\mb p}_{0}={\mb p}_{1}=\frac{1}{3}$&        &1.075 & 0.153 & 0.524 & $3.6\times10^{-6}$
&        & 0.551 & 0.565 & 1.676 & 0.856 & 0.547 & $1.4\times10^{-6}$\\
$p_{0}^{(opt)}$                 & 0.3326 &&&&& 0.3328 &&&&&&\\
$p_{-1}^{(opt)}=p_{1}^{(opt)}$  & 0.3326 &&&&& 0.3328 &&&&&&\\
${\mb p}_{-2}=\dots={\mb p}_{2}=\frac{1}{5}$&        &1.861 & 0.306 & 0.468 & $5.6\times10^{-6}$
&        & 1.730 & 0.058 & 0.206 & 0.252 & 0.491 & $5.0\times10^{-7}$\\
$p_{0}^{(opt)}$                 & 0.1994 &&&&& 0.1998 &&&&&&\\
$p_{-1}^{(opt)}=p_{1}^{(opt)}$  & 0.1994 &&&&& 0.1998 &&&&&&\\
$p_{-2}^{(opt)}=p_{2}^{(opt)}$  & 0.1993 &&&&& 0.1998 &&&&&&\\
${\mb p}_{-3}=\dots={\mb p}_{3}=\frac{1}{7}$&        &2.416 & 0.394 & 0.836 & $2.4\times10^{-3}$
&        & 2.205 & 0.451 & 0.739 & 0.260 & 2.112 & $6.7\times10^{-5}$\\
$p_{0}^{(opt)}$                 & 0.1370 &&&&& 0.1402 &&&&&&\\
$p_{-1}^{(opt)}=p_{1}^{(opt)}$  & 0.1276 &&&&& 0.1410 &&&&&&\\
$p_{-2}^{(opt)}=p_{2}^{(opt)}$  & 0.1605 &&&&& 0.1423 &&&&&&\\
$p_{-3}^{(opt)}=p_{3}^{(opt)}$  & 0.1248 &&&&& 0.1405 &&&&&&\\
${\mb p}_{-4}=\dots={\mb p}_{4}=\frac{1}{9}$&        &0.849 & 0.218 & 3.895 & $4.8\times10^{-3}$
&        & 2.436 & 0.267 & 1.627 & 0.402 & 0.949 & $1.3\times10^{-4}$\\
$p_{0}^{(opt)}$                 & 0.0869 &&&&& 0.1087 &&&&&&\\
$p_{-1}^{(opt)}=p_{1}^{(opt)}$  & 0.1105 &&&&& 0.1108 &&&&&&\\
$p_{-2}^{(opt)}=p_{2}^{(opt)}$  & 0.1034 &&&&& 0.1087 &&&&&&\\
$p_{-3}^{(opt)}=p_{3}^{(opt)}$  & 0.1323 &&&&& 0.1120 &&&&&&\\
$p_{-4}^{(opt)}=p_{4}^{(opt)}$  & 0.0802 &&&&& 0.1073 &&&&&&\\
\hline
\end{tabular}
\label{table1}
\end{table*}

\subsection{Two--pulse momentum distributions}
\label{2_pulse_distros}

For two pulses, the relevant parameters are the dimensionless pulse 
areas, $\alpha_{1}$ and $\alpha_{2}$, and the interval between the 
pulses, $\beta_{1}$, measured in units of the Talbot time.  We obtained 
least--squares--designed two--pulse sequences (which we shall refer to
as {\em optimal} distributions) for eight different specified momentum 
distributions.  These were four momentum magnitude distributions whose
specified probabilities are: 
$D_{mag}^{(2)}(1):p_{-1}=p_{1}=1/2$, $D_{mag}^{(2)}(2):p_{-2}=p_{2}=1/2$, 
$D_{mag}^{(2)}(3):p_{-3}=p_{3}=1/2$, and $D_{mag}^{(2)}(4):p_{-4}=p_{4}=1/2$.
As well as four momentum range distributions:
$D_{range}^{(2)}(1):p_{-1}=p_{0}=p_{1}=1/3$,
$D_{range}^{(2)}(2):p_{-2}=\dots=p_{2}=1/5$,
$D_{range}^{(2)}(1):p_{-3}=p_{0}=p_{3}=1/7$,
$D_{range}^{(2)}(1):p_{-4}=p_{0}=p_{4}=1/9$.
The results obtained for all of the optimal distributions in both the 
two--pulse and three--pulse cases are given in Table \ref{table1} and
in Figs.\ \ref{shaping_fig1} and \ref{shaping_fig2}.

Figure \ref{shaping_fig1}(a) shows graphs of two--pulse momentum magnitude
distributions ordered from left to right along the top row. Each graph exhibits
the probability for atoms to be in each momentum order versus of momentum 
orders $-5\le m\le 5$.  The values of the optimal two--pulse--sequence 
parameters and probabilities for each distribution are given in Table 
\ref{table1}. As can be seen from the figure, even with two pulses, a 
remarkably high degree of fidelity with the desired momentum distribution 
can be obtained with just two pulses.  For distribution $D_{mag}^{(2)}(1)$, 
43.3\% of the population is found in the $m=1$ and $m=-1$ for a total 
of 86.6\% populating momentum magnitude $2\hbar k$.  For distribution 
$D_{mag}^{(2)}(2)$ we find 42.7\% of the population in $m=2$ and $m=-2$ 
states and 35.7\% in $m=\pm 3$ (distribution $D_{mag}^{(2)}(3))$.  The 
maximum population achievable for two pulses degrades to 28.3\% for 
$m=\pm 4$.  We note that this is far better than is possible for a 
single pulse where the probability to populate states $\pm m$ is 
$J_{m}^{2}(\alpha)$.  In this case, the maximum values are 33.9\% 
($m=\pm 1$), 23.7\% ($m=\pm 2$), 18.9\% ($m=\pm 3$), and 16.0\% ($m=\pm 4$).
 
Two pulses do remarkably well in producing equal populations in a range 
of momentum states.  Plots of the optimal distributions determined by 
least--squares method are shown in Fig.\ \ref{shaping_fig1}(b) and column 
two of Table \ref{table1} gives the values of probabilities achieved for 
two pulses.  One can see that, for distribution $D_{range}^{(2)}(1)$, the 
populations in states $m=-1,0,1$ for distribution $D_{range}^{(2)}(1)$ 
are the approximately the same, $p_{0}^{(opt)} \approx p_{1}^{(opt)}=0.3326$, 
to four decimal places.  This is reasonably close to the specified value
of 0.3333.  In distribution $D_{range}^{(2)}(2)$, the populations in 
states $m=-2,-1,0,1,2$ the distribution across these states vary between 
0.1993 and 0.1994 which compares well with the specified value of 0.2.
For distribution, $D_{range}^{(2)}(3)$, while the states $m=-3,-2,-1,0,1,2,3$
contain more than 96\% of the total population, the probabilities vary
between 0.1248 and 0.1605 producing a relatively large variance around
the specified value of 1/7=0.1429.  The degradation of the fidelity can
especially be seen in distribution $D_{range}^{(2)}(4)$ both in the value 
of $(F_{LS}^{(2)})_{min}$ and in the deviation of the achievable 
probabilities from the specified probability as shown in Table \ref{table1}.  
For this distribution, the probabilites vary between 0.0802 and 0.1323
and deviate significantly from the specified value of 1/9=0.1111.

Upon closer inspection, we found that the optimal two--pulse
sequences had the common characteristic that there was a single dominant
pathway to each of the prescribed final momentum states.  Thus, for a fixed
final momentum state, this dominant pathway consisted of a direct jump from
the zero--momentum original condensate to the final momentum at the first pulse
followed by evolution between the pulses and no jump in momentum at the 
second pulse.  In this single--dominant--pathway picture, the total amplitude
is somewhat insensitive to the time between pulses since the coherent sum
can be (roughly) approximated with a single term.

\subsection{Three--pulse momentum distributions}
\label{3_pulse_distros}

For three pulses, the relevant parameters are the pulse areas, $\alpha_{1}$, 
$\alpha_{2}$, and $\alpha_{3}$, and the intervals between the pulses, 
$\beta_{1}$ and $\beta_{2}$.  We obtained optimal three--pulse sequences 
for the same set of eight specified momentum distributions as for two pulses.
As can be seen in Fig.\ \ref{shaping_fig2}, the fidelities achievable with
three pulses is better than with two pulses and is far superior to the 
single pulse case.

Figure \ref{shaping_fig2}(a) (top row) shows three--pulse momentum magnitude 
distributions $D_{mag}^{(3)}(1)$, $D_{mag}^{(3)}(2)$, $D_{mag}^{(3)}(3)$, 
and $D_{mag}^{(3)}(4)$ respectively from left to right.  While the bottom
row shows three--pulse momentum range distributions, $D_{range}^{(3)}(1)$, 
$D_{range}^{(3)}(2)$, $D_{range}^{(3)}(3)$, and $D_{range}^{(3)}(4)$ again
from left to right.  These plots show that the three--pulse optimal 
distributions faithfully reproduce the specified distributions better than
the two--pulse versions.  Comparisons of the fidelities for the two--pulse 
case versus the three--pulse given in Table \ref{table1} shows that three 
pulse does a better job at reproducing the specified distribution in every
case.

For some of the distributions there is little difference between two pulses
and three pulses because of the high fidelity of the two--pulse case.  One
example is $D_{range}^{(2)}(1)$ versus $D_{range}^{(3)}(1)$.  However, 
three pulses is clearly better for the $D_{range}^{(3)}(3)$ and
$D_{range}^{(3)}(4)$ distributions.

\section{Conclusion}
\label{conclusion}

In this paper we have shown that it is possible to create
clouds of coherent atoms with momentum--space distributions important for
applications in atom--interferometry with a sequence of only two or three
standing--wave laser pulses applied to a BEC.  We derived the momentum 
distribution for atoms in a BEC after $N$ short--time, standing--wave laser 
pulses were applied where the pulse areas and time intervals between the pulses
were variable.  This distribution was a generalization of the expression for 
two pulses whose validity has been verified 
experimentally~\cite{PhysRevLett.83.5407}.  We further described a method 
for designing pulse sequences that produce a specified momentum distribution 
of the condensate atoms.  We found that two kinds of distributions that have 
important applications as beam splitters can be produced with high fidelity 
with two or three pulses.  We also found that the optimal two--pulse sequences 
obtained could be understood in terms of a single--dominant--pathway picture.

The ability to produce coherent atom with engineered 
momentum--space distributions can now become a new tool for the design of
new atom interferometer schemes.  Methods for initial momentum--state 
selection or for producing multiple--beam splitters can now be designed.
For example, one could imagine a Bose--Einstein condensate created and 
confined on an atom chip to which could be applied a sequence of pulses
such as $D^{(3)}_{range}(2)$ so that the cloud is split into five equal
parts.  If such a condensate were confined by a harmonic trap potential,
these parts would eventually all come back together at once where they
could be split again.  This would produce multiple interference patterns 
that reflect the different phase evolutions along the different pathways.  
Such multi--particle interferometers could, in principle, implement quantum 
computations or be used for precision navigation applications, gradiometry,
or fundamental studies.

\appendix
\section{}
\label{app1}

This appendix presents the derivation of Eq.\ (\ref{1st_pulse}) which
expresses the condensate wave function just after the application of a 
short--time, standing--wave laser pulse in terms of the wave function
just before the pulse.  

We begin by transforming away the internal energies of the atom:
\begin{equation}
\psi_{k}\left({\bf r}_{a},t\right) \equiv 
e^{-iE_{k}t/\hbar}\phi_{k}\left({\bf r}_{a},t\right).\quad
k = g,e
\end{equation}
Under this transformation, Eqs.\ (\ref{mcgp_g}) and (\ref{mcgp_e})
become
\begin{eqnarray}
i\hbar\frac{\partial\phi_{g}}{\partial t} &=&
H_{0}^{(g)}\phi_{g}\left({\bf r}_{a},t\right)\nonumber\\
&+& 
\hbar\Omega_{0}\cos\left({\bf k}_{L}\cdot{\bf r}_{a}\right)
f\left(t\right)e^{-i\Delta t}
\phi_{e}\left({\bf r}_{a},t\right)\nonumber\\
&+& 
gN\left(
\left|\phi_{g}\left({\bf r}_{a},t\right)\right|^{2} +
\left|\phi_{e}\left({\bf r}_{a},t\right)\right|^{2}
\right)\phi_{g}\left({\bf r}_{a},t\right),\nonumber\\
\label{phi_g_app}
\end{eqnarray}
and
\begin{eqnarray}
i\hbar\frac{\partial\phi_{e}}{\partial t} &=&
H_{0}^{(e)}\phi_{e}\left({\bf r}_{a},t\right)\nonumber\\
&+& 
\hbar\Omega_{0}\cos\left({\bf k}_{L}\cdot{\bf r}_{a}\right)
f\left(t\right)e^{i\Delta t}
\phi_{g}\left({\bf r}_{a},t\right)\nonumber\\
&+& 
gN\left(
\left|\phi_{g}\left({\bf r}_{a},t\right)\right|^{2} +
\left|\phi_{e}\left({\bf r}_{a},t\right)\right|^{2}
\right)\phi_{e}\left({\bf r}_{a},t\right),\nonumber\\
\label{phi_e_app}
\end{eqnarray}
where we have made the rotating--wave approximation (RWA).
The RWA consists of neglecting the exponentials that oscillate as 
$\exp\left[\pm i\left(\omega_{0} + \omega_{L}\right)t\right]$ relative to 
$\exp\left(\pm i\Delta t\right)$ where $\hbar\omega_{0}=E_{e}-E_{g}$.
This approximation holds because $\omega_{L}$ is typically six orders 
of magnitude larger than $\Delta$ for optical frequencies when
the light is close to resonance.

Since we assume $\Omega_{0} \ll \Delta$, there is never 
very much population in the upper state.  Thus we can neglect
the nonlinear term and kinetic plus trap potential energy 
terms in Eq.\ (\ref{phi_e_app}) and can neglect $\phi_{e}$ in 
the nonlinear term of Eq.\ (\ref{phi_g_app}).  This gives
\begin{eqnarray}
i\hbar\frac{\partial\phi_{g}}{\partial t} &\approx&
H_{0}^{(g)}\phi_{g}\left({\bf r}_{a},t\right) +
gN
\left|\phi_{g}\left({\bf r}_{a},t\right)\right|^{2}
\phi_{g}\left({\bf r}_{a},t\right)\nonumber\\
&+& 
\hbar\Omega_{0}\cos\left({\bf k}_{L}\cdot{\bf r}_{a}\right)
f\left(t\right)e^{-i\Delta t}
\phi_{e}\left({\bf r}_{a},t\right)\nonumber\\
\label{phig_a1_app}
\end{eqnarray}
\begin{eqnarray}
i\hbar\frac{\partial\phi_{e}}{\partial t} &\approx&
\hbar\Omega_{0}\cos\left({\bf k}_{L}\cdot{\bf r}_{a}\right)
f\left(t\right)e^{i\Delta t}
\phi_{g}\left({\bf r}_{a},t\right)
\label{phie_a1_app}
\end{eqnarray}
Finally, note that the first two terms on the right--hand--side
of Eq.\ (\ref{phig_a1_app}) approximately satisfy the static GP
equation and thus together equal $\mu\phi_{g}$.  Replacing
those two terms gives:
\begin{eqnarray}
i\hbar\frac{\partial\phi_{g}}{\partial t} &\approx&
\mu\phi_{g}\left({\bf r}_{a},t\right)\nonumber\\
&+& 
\hbar\Omega_{0}\cos\left({\bf k}_{L}\cdot{\bf r}_{a}\right)
f\left(t\right)e^{-i\Delta t}
\phi_{e}\left({\bf r}_{a},t\right)\nonumber\\
\label{phig_a2_app}
\end{eqnarray}
This term can be transformed away but it is unecessary since
we have assumed that $\mu\delta t/\hbar \ll 1$.  Neglecting
this term finally gives
\begin{eqnarray}
i\hbar\frac{\partial\phi_{g}}{\partial t} &\approx&
\hbar\Omega_{0}\cos\left({\bf k}_{L}\cdot{\bf r}_{a}\right)
f\left(t\right)e^{-i\Delta t}
\phi_{e}\left({\bf r}_{a},t\right)
\label{phig_fin_app}
\end{eqnarray}
\begin{eqnarray}
i\hbar\frac{\partial\phi_{e}}{\partial t} &\approx&
\hbar\Omega_{0}\cos\left({\bf k}_{L}\cdot{\bf r}_{a}\right)
f\left(t\right)e^{i\Delta t}
\phi_{g}\left({\bf r}_{a},t\right)
\label{phie_fin_app}
\end{eqnarray}
Under the ``sudden approximation,'' these equations can be
easily solved.  That is, we assume that the turn--on of the
pulse is fast enough that the atom remains in its initial
state until the pulse is fully on.

In order to express the solution in terms of the area of a
single pulse, we transform the solutions as
\begin{equation}
\left(
\begin{array}{c}
\phi_{g}\left({\bf r}_{a},t\right)\\
\phi_{e}\left({\bf r}_{a},t\right)
\end{array}
\right) =
\left(
\begin{array}{cc}
e^{-i\Delta t/2} & 0\\
0 & e^{i\Delta t/2}
\end{array}
\right)
\left(
\begin{array}{c}
\bar{\phi}_{g}\left({\bf r}_{a},t\right)\\
\bar{\phi}_{e}\left({\bf r}_{a},t\right)
\end{array}
\right)
\end{equation}
This yields the following equations for $\bar{\phi}_{g,e}$:
\begin{eqnarray}
i\hbar\frac{\partial\bar{\phi}_{g}}{\partial t} &=&
-\frac{1}{2}\hbar\Delta
\bar{\phi}_{g}\left({\bf r}_{a},t\right) + 
V\left({\bf r}_{a},t\right)
\bar{\phi}_{e}\left({\bf r}_{a},t\right)\nonumber\\
i\hbar\frac{\partial\bar{\phi}_{e}}{\partial t} &=&
\ \ \,\frac{1}{2}\hbar\Delta
\bar{\phi}_{e}\left({\bf r}_{a},t\right) + 
V\left({\bf r}_{a},t\right)
\bar{\phi}_{g}\left({\bf r}_{a},t\right),
\label{fin_eqs_app}
\end{eqnarray}
where
\begin{equation}
V\left({\bf r}_{a},t\right) = \hbar\Omega_{0}f\left(t\right)
\cos\left({\bf k}_{L}\cdot{\bf r}_{a}\right).
\end{equation}
Defining
\begin{equation}
\bar{\Phi}\left({\bf r}_{a},t\right) \equiv
\left(
\begin{array}{c}
\bar{\phi}_{g}\left({\bf r}_{a},t\right)\\
\bar{\phi}_{e}\left({\bf r}_{a},t\right)
\end{array}
\right),
\end{equation}
and
\begin{equation}
{\cal H}\left({\bf r}_{a},t\right) \equiv
\left(
\begin{array}{cc}
-\frac{1}{2}\hbar\Delta & V\left({\bf r}_{a},t\right)\\
V\left({\bf r}_{a},t\right) & \frac{1}{2}\hbar\Delta
\end{array}
\right)
\label{h_def_app}
\end{equation}
we may formally express Eqs.\ (\ref{fin_eqs_app}) as
\begin{equation}
i\hbar\frac{\partial\bar{\Phi}}{\partial t} = 
{\cal H}\left({\bf r}_{a},t\right)
\bar{\Phi}\left({\bf r}_{a},t\right).
\label{formal_eq_app}
\end{equation}
Since ${\cal H}\left({\bf r}_{a},t\right)$ is constant
during the pulse we obtain a relationship between 
$\bar{\Phi}$ before and after the pulse
\begin{equation}
\bar{\Phi}\left({\bf r}_{a},t_{0}+\delta t/2\right) = 
e^{-iM\delta t}
\bar{\Phi}\left({\bf r}_{a},t_{0}-\delta t/2\right),
\label{solution_app}
\end{equation}
where
\begin{equation}
M = 
\left(
\begin{array}{cc}
-\frac{1}{2}\Delta & 
\Omega_{0}\cos\left({\bf k}_{L}\cdot{\bf r}_{a}\right)\\
\Omega_{0}\cos\left({\bf k}_{L}\cdot{\bf r}_{a}\right) & 
\frac{1}{2}\Delta
\end{array}
\right).
\end{equation}
This matrix can be easily exponentiated using its eigenvalues,
\begin{equation}
\lambda_{\pm} =
\pm\left[
\left(\scriptstyle{\frac{1}{2}}\displaystyle\Delta\right)^{2} +
\Omega_{0}^{2}\cos^{2}\left({\bf k}_{L}\cdot{\bf r}_{a}\right)
\right]^{1/2} \equiv \pm\lambda,
\end{equation}
and eigenvectors
\begin{equation}
|+\lambda\rangle = 
\left(
\begin{array}{c}
\sin\left(\theta/2\right)\\
\cos\left(\theta/2\right)
\end{array}
\right)
\quad
|-\lambda\rangle = 
\left(
\begin{array}{c}
\cos\left(\theta/2\right)\\
-\sin\left(\theta/2\right)
\end{array}
\right),
\end{equation}
where $\theta$ is defined by
\begin{equation}
\sin\left(\theta\right) = 
\frac{\Omega_{0}\cos\left({\bf k}_{L}\cdot{\bf r}_{a}\right)}
{\lambda},\quad
\cos\left(\theta\right) = 
\frac{\scriptstyle{\frac{1}{2}}\displaystyle\Delta}{\lambda}.
\end{equation}
The diagonalization matrix, $U^{\dag}$, where
\begin{equation}
M_{D} = UMU^{\dag} = 
\left(
\begin{array}{cc}
\lambda & 0\\
0 & -\lambda
\end{array}
\right),
\end{equation}
is given by
\begin{equation}
U^{\dag} = 
\left(
\begin{array}{cc}
\sin\left(\theta/2\right) & \cos\left(\theta/2\right)\\
\cos\left(\theta/2\right) &-\sin\left(\theta/2\right)
\end{array}
\right)
\end{equation}
The exponentiated matrix is, therefore
\begin{equation}
e^{-iM\delta t} = 
\left(
\begin{array}{cc}
s^{2}e^{-i\lambda\delta t} + c^{2}e^{i\lambda\delta t}
& -sc\left(e^{i\lambda\delta t}-e^{-i\lambda\delta t}\right)\\
-sc\left(e^{i\lambda\delta t}-e^{-i\lambda\delta t}\right)
& c^{2}e^{-i\lambda\delta t} + s^{2}e^{i\lambda\delta t}
\end{array}
\right)\label{eq1_app}\nonumber
\end{equation}
where $s \equiv \sin\left(\theta/2\right)$ and 
$c \equiv \cos\left(\theta/2\right)$.

Thus we can write a relationship between $\bar{\phi}_{g}$ 
before and after the pulse.  From Eqs.\ (\ref{solution_app})
and (\ref{eq1_app}) we have
\begin{equation}
\begin{array}{cc}
\bar{\phi}_{g}\left({\bf r}_{a},t_{+}\right) = &
\left(
s^{2}e^{-i\lambda\delta t} + c^{2}e^{i\lambda\delta t}
\right)
\bar{\phi}_{g}\left({\bf r}_{a},t_{-}\right)\\
&
-\,sc\left(e^{i\lambda\delta t}-e^{-i\lambda\delta t}\right)
\bar{\phi}_{e}\left({\bf r}_{a},t_{-}\right)
\end{array}
\label{end_eq_app}
\end{equation}
where $t_{\pm}\equiv t_{0}\pm\delta t/2$.

We can simplify the above expression by invoking the 
approximation that the detuning from resonance is much
larger that the single--photon Rabi frequency, 
$\Omega_{0} \ll \Delta$.  In this case, we have
\begin{eqnarray}
\lambda &\approx& 
\frac{1}{2}\Delta +
\left(
\frac{\Omega_{0}^{2}}{\Delta}
\right)
\cos^{2}\left({\bf k}_{L}\cdot{\bf r}_{a}\right)\nonumber\\
&=&
\frac{1}{2}\Delta +
\frac{1}{2}
\left(
\frac{\Omega_{0}^{2}}{\Delta}
\right)
\left(
1 +
\cos\left(2{\bf k}_{L}\cdot{\bf r}_{a}\right)
\right)
\end{eqnarray}
Furthermore,
\begin{eqnarray}
\cos\left(\theta\right) 
&=&
\frac{\frac{1}{2}\Delta}{\lambda}\nonumber\\
&=&
\frac{\frac{1}{2}\Delta}
{
\left(
\left(
\frac{1}{2}\Delta
\right)^{2} +
\Omega_{0}^{2}
\cos^{2}\left({\bf k}_{L}\cdot{\bf r}_{a}\right)
\right)^{1/2}
}\nonumber\\
&\approx&
1 - \frac{1}{2}
\left(
\frac{\Omega_{0}}{\frac{1}{2}\Delta}
\right)^{2}
\cos^{2}\left({\bf k}_{L}\cdot{\bf r}_{a}\right).
\end{eqnarray}
The factors $s^{2}$ and $c^{2}$ are, to second order
in $\Omega_{0}/\Delta$, approximately
\begin{eqnarray}
s^{2} &\equiv& \sin^{2}\left(\theta/2\right)\nonumber\\
&=&
\frac{1}{2}
\left(
1 - \cos\left(\theta\right)
\right) =
\frac{1}{2}
\left(
1 - \frac{\frac{1}{2}\Delta}{\lambda}
\right)
\nonumber\\
s^{2}
&\approx&
\left(\frac{\Omega_{0}}{\Delta}\right)^{2}
\cos^{2}\left({\bf k}_{L}\cdot{\bf r}_{a}\right)\\
c^{2} &\equiv& \cos^{2}\left(\theta/2\right)\nonumber\\
&=&
\frac{1}{2}
\left(
1 + \cos\left(\theta\right)
\right) =
\frac{1}{2}
\left(
1 + \frac{\frac{1}{2}\Delta}{\lambda}
\right)
\nonumber\\
c^{2}
&\approx&
1 - \left(\frac{\Omega_{0}}{\Delta}\right)^{2}
\cos^{2}\left({\bf k}_{L}\cdot{\bf r}_{a}\right).
\end{eqnarray}
Given the above it should be an excellent approximation 
to set $s\approx 0$ and $c\approx 1$.  Hence 
\begin{equation}
\bar{\phi}_{g}\left({\bf r}_{a},t_{+}\right) \approx
e^{
i(\frac{1}{2}\Delta + \Omega_{2})\delta t +
i\Omega_{2}\delta t\cos\left(2{\bf k}_{L}\cdot{\bf r}_{a}\right)
}
\bar{\phi}_{g}\left({\bf r}_{a},t_{-}\right),
\end{equation}
where
\begin{equation}
\Omega_{2} \equiv \frac{\Omega_{0}^{2}}{2\Delta}
\end{equation}
is the two--photon Rabi frequency.  We can express the 
above relationship in terms of the original condensate
wave function as
\begin{eqnarray}
\phi_{g}\left({\bf r}_{a},t_{+}\right) &\approx&
e^{i\Omega_{2}\delta t}\nonumber\\
&&
\times
e^{i\Omega_{2}\delta t\cos\left(2{\bf k}_{L}\cdot{\bf r}_{a}\right)}
\phi_{g}\left({\bf r}_{a},t_{-}\right).
\label{1st_pulse_app}
\end{eqnarray}
This is the effect of a single standing--wave pulse on the 
condensate wave function.

\begin{acknowledgments}
We acknowledge helpful discussions with L.\ Deng, E.W.\ Hagley, 
I.\ Spielman, A.\ Cassidy, G.\ Campbell, and W.D.\ Phillips.  This work 
was supported in part by the U.S.\ National Science Foundation grant 
PHY--0758111, the Physics Frontiers Center grant PHY--0822671 and by 
the National Institute of Standards and Technology.
\end{acknowledgments}

\bibliography{momentum_shaping_pra_revised}{}

\end{document}